\documentclass[fp,twocolumn]{jpsj3}
\usepackage{txfonts,bm,amsmath,amssymb,url,float,cases,braket,xcolor,graphicx}
\begin{document}

\title{
Total Moment Sum Rule for Magnets in the Vicinity of Quantum Critical Point
}

\author{
Masashige Matsumoto\thanks{E-mail address: matsumoto.masashige@shizuoka.ac.jp}
}

\inst{
Department of Physics, Shizuoka University, Shizuoka 422-8529, Japan
}

\recdate{September  28, 2020}

\abst{
It is known that the longitudinal and transverse excitation modes can exist
in the vicinity of a quantum critical point in the ordered phase of quantum magnetic systems.
The total moment sum rule for such systems is derived on the basis of the extended spin-wave theory,
where both longitudinal and transverse magnetic excitations are taken into account.
The sum rule is resolved into elastic, one-magnon, and two-magnon components.
The formulation is applicable to spin systems with the longitudinal mode,
such as $S=1$ systems with single-ion anisotropy of easy-plane type and spin dimer systems.
The result helps us analyze and understand measured data of inelastic neutron scattering.
}

\maketitle


\renewcommand{\H}{{\mathcal H}}

\newcommand{\bH}{{\bm H}}
\newcommand{\bQ}{{\bm Q}}
\newcommand{\bq}{{\bm q}}
\newcommand{\bk}{{\bm k}}
\newcommand{\br}{{\bm r}}
\newcommand{\bS}{{\bm S}}
\newcommand{\be}{{\bm e}}

\newcommand{\rL}{{\rm L}}
\newcommand{\rT}{{\rm T}}

\newcommand{\Cr}{{Cr$_2$WO$_6$}}
\newcommand{\CrTe}{{Cr$_2$TeO$_6$}}
\newcommand{\Ba}{{Ba$_2$CoGe$_2$O$_7$}}
\newcommand{\Tl}{{TlCuCl$_3$}}
\newcommand{\Cu}{{Cu(NO$_3$)$_2\cdot$2.5D$_2$O}}
\newcommand{\Ca}{{Ca$_2$RuO$_4$}}
\newcommand{\DLCB}{C$_9$H$_{18}$N$_2$CuBr$_4$}
\newcommand{\Cs}{{CsFeCl$_3$}}
\newcommand{\CsNi}{{CsNiCl$_3$}}


\section{Introduction}

The total moment sum rule is a kind of conservation law in dynamical spin correlation function.
When we integrate the correlation function over the frequency $\omega$, momentum $\bq$,
and $x$, $y$, and $z$ components of spin $S$ operator, we obtain
\cite{Zhu-2005}
\begin{align}
\sum_{\alpha=x,y,z} \frac{1}{N} \sum_\bq \int d\omega S^{\alpha\alpha}(\bq,\omega) = S(S+1).
\end{align}
Here, $N$ is the number of spins in a sample.
Since the dynamical spin correlation function is related to the intensity of inelastic neutron scattering,
the sum rule can be used to analyze the measured experimental date of inelastic neutron scattering.
Based on the conventional spin-wave theory, the integrated correlation function can be resolved
into elastic and inelastic (one-magnon, two-magnon, and multi-magnon) components.
\cite{Huberman-2005}
This was applied to estimate the intensity of the two-magnon scattering relative to that for the one-magnon
in $S=5/2$ Heisenberg antiferromagnet on a square lattice.
\cite{Huberman-2005}
It was also applied to a spin ladder system to derive an expression for the dynamic spin-correlation function.
\cite{Hammar-1998}
Thus, the total moment sum rule helps us analyze and understand the measured magnetic excitations by neutron scattering.

On the other hand, the conventional spin-wave theory fails to describe magnetic excitations
in spin systems that show a quantum phase transition,
such as in spin dimer systems where both transverse and longitudinal excitations exist
in the vicinity of the quantum critical point in the ordered state.
Recently, inelastic neutron scattering experiments were performed in a quantum spin system \Cr.
\cite{Zhu-2019}
The crystal structure shows that two Cr$^{3+}$ ions are strongly coupled and form a spin dimer by S = 3/2 spins.
At low temperatures, a magnetically ordered state is stabilized.
\cite{Zhu-2014}
They found unexpected high-energy magnetic excitations with weak intensity above the conventional spin-wave excitation mode.
\cite{Zhu-2019}
The conventional spin-wave theory accounts for the low-energy transverse spin-wave mode,
however, it is difficult to explain the origin of the additional high-energy excitations.
These measurements were performed with polycrystalline samples,
since it was difficult to synthesize single crystals.

The extended spin-wave theory, or the generalized Holstein-Primakoff theory,
is a powerful tool for investigating magnetic excitations in the vicinity of the quantum critical point,
such as in spin dimer systems.
\cite{Papanicolaou-1984,Onufrieva-1985,Joshi-1999,Shiina-2003,Shiina-2004}
The formulation is equivalent to the harmonic bond-operator theory
and can describe both transverse and longitudinal magnetic excitations.
\cite{Sachdev-1990,Sommer-2001,Matsumoto-2002,Matsumoto-2004}
The extended spin-wave theory is also equivalent to the multi-boson spin-wave theory
introduced to study magnetic excitations in S = 3/2 spin system with a strong single-ion anisotropy of an easy-plane type,
such as in \Ba.
\cite{Penc-2012,Rhomhanyi-2012,Akaki-2017}
The characteristic point is that plural bosons are introduced for each spin multiplet of the local spin states.
This enables us to describe both the low-energy transverse and high-energy longitudinal modes.

Recently, the longitudinal excitation mode is termed as Higgs amplitude mode in quantum magnets
in the similarity to the Higgs particle in high-energy physics.
\cite{Sachdev-2011,Pekker-2015}
The observation of the longitudinal mode (L-mode) is one of the hot topics in condensed matter physics.
Since \Cr~is a spin dimer system, we can expect the L-mode, as in \Tl.
\cite{Ruegg-2008}
In the presence of an easy-axis anisotropy in the exchange interactions,
the L-mode was also observed by spin-polarized inelastic neutron scattering measurements
in a two-dimensional spin-1/2 coupled two-leg spin ladder antiferromagnet.
\cite{Hong-2017}
The L-mode was also pointed out in $S=1$ systems with single-ion anisotropy of easy-plane type,
\cite{Affleck-1989,Plumer-1992,Affleck-1992,Matsumoto-2007}
and it was observed in \Ca.
\cite{Jain-2017}
In \CsNi~under the atmospheric pressure, the T- and L-modes are hybridized by the noncollinear magnetic structure.
\cite{Morra-1988,Tun-1990,Kakurai-1991,Affleck-1992}
In \Cs, this point was elucidated by measuring the evolution of inelastic neutron spectra
under controlling pressure to cross the quantum critical point.
\cite{Hayashida-2019,Matsumoto-2020}

About the high-energy excitations observed in \Cr, there are two main possibilities for the interpretation.
One is an L-mode in one-magnon process.
The other is a transverse mode (T-mode) in two-magnon process.
Theoretically, it is difficult to calculate the intensity for two-magnon process in polycrystalline samples.
To judge the origin of the high-energy excitations in \Cr, we can use the total moment sum rule for spin dimer systems,
where the integrated intensity is resolved into one magnon L-mode and two-magnon T-mode.

Since there is no theory of the total moment sum rule for spin dimer systems so far,
we derive the sum rule based on the extended spin-wave theory.
The sum rule should be different from the conventional spin systems.
We apply the result to two dimer systems \Cu
\cite{Tennant-2003}
and \Cr,
\cite{Hong-2017}
since the former and the latter are in disordered and ordered phases at low temperatures, respectively,
and the difference between the two cases becomes prominent.

This paper is organized as follows.
In Sect. 2, we briefly summarize the total moment sum rule based on the conventional spin-wave theory.
In Sect. 3, we introduce the extended spin-wave theory.
We apply the theory to an $S=1$ quantum spin system with a single-ion anisotropy of easy-plane type in Sect. 4.
In Sect. 5, the total moment sum rule in $S=1/2$ spin dimer systems is derived.
The result is extended to an $S=3/2$ dimer case and applied to \Cr.
The last section gives a summary of the paper.

\section{Total Moment Sum Rule}

Let us explain details of the total moment sum rule and resolve it into elastic and inelastic components
based on the conventional spin-wave theory.
\cite{Huberman-2005}
This helps us understand how to apply the extended spin-wave theory to the sum rule discussed in the next section.
The dynamical spin correlation function is defined by
\begin{align}
S^{\alpha\alpha}(\bq,\omega) = \frac{1}{2\pi} \int dt e^{-i\omega t} \langle S^\alpha(\bq) S^\alpha(-\bq,t) \rangle.
\label{eqn:Sqw}
\end{align}
Here, $S^\alpha(\bq)$ represents the Fourier transformed $\alpha(=x,y,z)$ component the spin operator.
It is defined by
\begin{align}
S^\alpha(\bq) = \frac{1}{\sqrt{N}} \sum_i e^{-i \bq\cdot \br_i} S^\alpha_i,
\end{align}
where $S^\alpha_i$ is the spin operator on the $i$th site at position $\br_i$.
$N$ represents the total number of spins.
In Eq. (\ref{eqn:Sqw}), $S^\alpha(-\bq,t)$ represents the Heisenberg representation of $S^\alpha(-\bq)$.
Integrating the dynamical spin correlation function over $\bq$ and $\omega$ and summing up $\alpha=x,y,z$,
we obtain the following total moment sum rule:
\begin{align}
&\sum_{\alpha=x,y,z} \frac{1}{N} \sum_\bq \int d\omega S^{\alpha\alpha}(\bq,\omega) \cr
&= \sum_{\alpha=x,y,z}  \frac{1}{N} \sum_{i,j} \int dt  \langle S^\alpha_i S^\alpha_j(t) \rangle
\frac{1}{2\pi} \int d\omega e^{-i \omega t} \frac{1}{N} \sum_\bq e^{i\bq\cdot(\br_j-\br_i)} \cr
&=\frac{1}{N} \sum_i \langle ( S_i^x S_i^x + S_i^y S_i^y + S_i^z S_i^z ) \rangle \cr
&=S(S+1).
\label{eqn:sum}
\end{align}
This result represents that the integrated dynamical spin correlation function is fixed to the specified value.
In this section, we resolve the total moment sum rule into elastic and inelastic parts within the conventional spin-wave theory.

\subsection{Ordered moment}

We consider the following Hamiltonian for antiferromagnets on a simple cubic lattice:
\begin{align}
\H = J \sum_{\langle i,j\rangle} \bS_i \cdot \bS_j.
\label{eqn:H}
\end{align}
Here, $\bS_i$ and $\bS_j$ are spin operators at the $i$th and $i$th sites, respectively.
$J$ is an exchange interaction parameter.
The summation $\sum_{\langle i,j\rangle}$ is taken over the nearest neighbor sites.
We consider an antiferromagnetic (AF) case with $J>0$.
In the N\'{e}el ordered state, we take the $z$-axis along the ordered moment.

As shown in the Appendix, the expectation value of the ordered moment on the A (up) sublattice is given by
\begin{align}
\langle S_i^z \rangle = S - \langle a_i^\dagger a_i \rangle.
\end{align}
Here, $a_i^\dagger$ and $a_i$ are creation and annihilation Bose operators at the $i$th site
introduced in the Holstein-Primakoff transformation.
Throughout this paper, we take the expectation value by the ground state in the low temperature limit $T\rightarrow 0$.
The expectation value of the population of the boson is calculated as
\begin{align}
n &= \langle a_i^\dagger a_i \rangle
= \frac{1}{N/2} \sum_\bk \langle a_\bk^\dagger a_\bk \rangle \cr
&= \frac{1}{N/2} \sum_\bk \left( u_\bk^2 \langle \alpha_\bk^\dagger \alpha_\bk \rangle
                               + v_\bk^2 \langle \beta_\bk \beta_\bk^\dagger \rangle \right) \cr
&= \frac{1}{N/2} \sum_\bk v_\bk^2.
\label{eqn:n}
\end{align}
Here, $N/2$ represents the number of spin site on the A sublattice,
and used $\braket{\alpha_\bk^\dagger \alpha_\bk}=0$ at low temperatures.
$u_\bk$ and $v_\bk$ are coefficients of the Bogoliubov transformation [see Eq. (\ref{eqn:Bogoliubov})].
We have used the Fourier transformation given by Eq. (\ref{eqn:Fourier}).
Notice that the expectation value does not depend on the spin site.
On the B (down) sublattice, we obtain the same result.
The ordered moment is then expressed as
\begin{align}
|\langle S_i^z \rangle|
&= S - n.
\label{eqn:Sz-av}
\end{align}
Here, $n$ represents the reduction of the moment by the quantum effect
on the basis of the conventional spin-wave theory.

\subsection{$z$ component}

As in Eq. (\ref{eqn:sum}), the $z$ component of the integrated correlation function is expressed by
\begin{align}
&\frac{1}{N} \sum_\bq \int d\omega S^{zz}(\bq,\omega) = \frac{1}{N} \sum_i \langle S^z_i S^z_i \rangle \cr
&~~~~~~~~~~~~~~~~~~~~~~~~
= \frac{1}{N} \sum_i
\int d\omega
\frac{1}{2\pi}
\int dt  e^{-i \omega t} \langle S^z_i S^z_i(t) \rangle.
\label{eqn:Szz}
\end{align}
In the second line of Eq. (\ref{eqn:Szz}), notice that we first perform the integral over $t$
in order to discuss elastic and inelastic components of the dynamical spin correlation function.
On the A sublattice, we substitute Eq. (\ref{eqn:Sa}) into Eq. (\ref{eqn:Szz}) and obtain
\begin{align}
&\langle S^z_i S^z_i(t) \rangle
= \langle ( S - a_i^\dagger a_i ) [ S - a_i^\dagger(t) a_i(t) ] \rangle \cr
&= S^2
- S \langle a_i^\dagger a_i \rangle
- S \langle a_i^\dagger(t) a_i(t) \rangle
+ \langle a_i^\dagger a_i a_i^\dagger(t) a_i(t) \rangle.
\label{eqn:Szz2}
\end{align}
The first two terms have no time dependence.
When we perform the integral over $t$ in Eq. (\ref{eqn:Szz}),
this leads to $1/(2\pi) \int dt e^{-i\omega t} =\delta(\omega)$ and $\omega$ must be zero.
Therefore, they are elastic component in the dynamical spin correlation function $S^{zz}(\bq,\omega)$.
As in Eq. (\ref{eqn:n}), the third term in Eq. (\ref{eqn:Szz2}) is calculated as
\begin{align}
&\langle a_i(t)^\dagger a_i(t) \rangle \cr
&= \frac{1}{N/2} \sum_\bk \left( u_\bk^2 \langle \alpha_\bk^\dagger e^{i E_\bk t} \alpha_\bk e^{-i E_\bk t} \rangle
                                                         + v_\bk^2 \langle \beta_\bk e^{-i E_\bk t} \beta_\bk^\dagger e^{i E_\bk t} \rangle \right) \cr
&= \frac{1}{N/2} \sum_\bk v_\bk^2
= n.
\end{align}
Since the time dependence cancels out, the third term in Eq. (\ref{eqn:Szz2}) is also elastic component.
The last term in Eq. (\ref{eqn:Szz2}) is calculated as
\begin{align}
&\langle a_i^\dagger a_i a_i^\dagger(t) a_i(t) \rangle \cr
&= \left( \frac{1}{N/2} \right)^2 \sum_{\bk_1,\bk_2,\bk_3,\bk_4} v_{\bk_1} v_{\bk_2} v_{\bk_3} v_{\bk_4}
\langle \beta_{-\bk_1} \beta_{-\bk_2}^\dagger \rangle \langle \beta_{-\bk_3}(t) \beta_{-\bk_4}^\dagger(t) \rangle \cr
&~~~
+ \left( \frac{1}{N/2} \right)^2 \sum_{\bk_1,\bk_2,\bk_3,\bk_4} v_{\bk_1} u_{\bk_2} u_{\bk_3} v_{\bk_4}
\langle \beta_{-\bk_1} \beta_{-\bk_4}^\dagger(t) \rangle \langle \alpha_{\bk_2} \alpha_{\bk_3}^\dagger(t) \rangle \cr
&= \left( \frac{1}{N/2} \sum_{\bk_1} v_{\bk_1}^2 \right) \left( \frac{1}{N/2} \sum_{\bk_3} v_{\bk_3}^2 \right) \cr
&~~~
+ \left( \frac{1}{N/2} \sum_{\bk_1} v_{\bk_1}^2 e^{i E_{\bk_1} t} \right) \left( \frac{1}{N/2} \sum_{\bk_2} u_{\bk_2}^2 e^{i E_{\bk_2} t} \right).
\label{eqn:a4}
\end{align}
Here, the first term has no time dependence and is elastic component.
On the other hand, the second term has the time dependence $e^{i(E_{\bk_1}+E_{\bk_2})t}$
and leads to
$1/(2\pi) \int dt e^{-i(\omega - E_{\bk_1}-E_{\bk_2})t} = \delta(\omega - E_{\bk_1}-E_{\bk_2})$ in Eq. (\ref{eqn:Szz}).
Since this represents two-magnon excitation ($\omega=E_{\bk_1}+E_{\bk_2}$),
the second term in Eq. (\ref{eqn:a4}) is inelastic component in the dynamical spin correlation function $S^{zz}(\bq,\omega)$.
In the above discussion, we first performed the integral over $t$
and we understand how to resolve the dynamical spin correlation function into elastic and inelastic components.

Next, we first perform the integral over $\omega$ in Eq. (\ref{eqn:Szz}).
In this case, we obtain the delta function $\delta(t)$ and obtain the first line in Eq. (\ref{eqn:Szz}).
This means that we can put $t=0$ in evaluating $\braket{a_i^\dagger a_i a_i^\dagger(t) a_i(t)}$.
For $t=0$, Eq. (\ref{eqn:a4}) reduces to
\begin{align}
\braket{a_i^\dagger a_i a_i^\dagger(t) a_i(t)}|_{t\rightarrow 0} = n^2 + n( 1 + n ).
\label{eqn:t=0}
\end{align}
Here, we used Eq. (\ref{eqn:n}) and the relation $u_\bk^2 = 1 + v_\bk^2$ [see Eq. (\ref{eqn:uv})].
As discussed above, the $n^2$ and $n(1+n)$ terms are elastic and inelastic components, respectively.
The result in Eq. (\ref{eqn:t=0}) indicates that the expectation value can be simply separated as
\begin{align}
\langle a_i^\dagger a_i a_i^\dagger(t) a_i(t) \rangle
&= \langle a_i^\dagger a_i \rangle_{\rm elastic}~\langle a_i^\dagger(t) a_i(t) \rangle_{\rm elastic} \cr
&~~~
+ \langle a_i^\dagger a_i(t) \rangle_{\rm inelastic}~\langle a_i a_i^\dagger(t) \rangle_{\rm inelastic}.
\label{eqn:contraction}
\end{align}
In the practical calculation, we can put $t=0$ in Eq. (\ref{eqn:contraction}) after the integral over $\omega$.
We use this representation in the following discussions.

On the B sublattice, we obtain the same result.
Thus, the integrated correlation function of the $z$ component can be resolved into elastic and inelastic components.
In Eq. (\ref{eqn:Szz}), they are expressed as
\begin{equation}
\begin{aligned}
&\langle S^z_i S^z_i \rangle_{\rm elastic} = S^2 - 2Sn + n^2 = ( S - n )^2, \cr
&\langle S^z_i S^z_i \rangle_{\rm inelastic} = n(1+n).
\end{aligned}
\end{equation}

\subsection{$x$ and $y$ components}

The $x$ and $y$ components of the integrated correlation function is expressed as
\begin{align}
\frac{1}{N} \sum_\bq \int d\omega \left[ S^{xx}(\bq,\omega) + S^{yy}(\bq,\omega) \right] \cr
= \frac{1}{N} \sum_i \frac{1}{2} \langle ( S_i^+ S_i^- + S_i^- S_i^+ ) \rangle,
\label{eqn:moment-xy}
\end{align}
with $S_i^\pm = S_i^x \pm i S_i^y$.
Here, the integral over $\omega$ was performed and we put $t=0$.
Therefore, the operator in Eq. (\ref{eqn:moment-xy}) is understood as $[ S_i^+ S_i^-(t) + S_i^- S_i^+(t) ]|_{t\rightarrow 0}$.
As the first term in Eq. (\ref{eqn:a4}), elastic component is calculated by taking the following contraction:
$\langle S_i^+ \rangle \langle S_i^- \rangle + \langle S_i^- \rangle \langle S_i^+ \rangle$.
Since $\langle S_i^\pm \rangle = 0$, there no elastic component in Eq. (\ref{eqn:moment-xy}).

We next consider inelastic component.
Using Eq. (\ref{eqn:Sa}), we can express
\begin{align}
\frac{1}{2} ( S_i^+ S_i^- + S_i^- S_i^+ ) = S + 2S a_i^\dagger a_i - a_i^\dagger a_i a_i^\dagger a_i
\label{eqn:xy-component}
\end{align}
on the A sublattice.
In this calculation, we notice that not only one-magnon but also multi-magnon processes appear.
As in the $z$ component case, we can calculate the inelastic component as
\begin{align}
\frac{1}{2} \langle ( S_i^+ S_i^- + S_i^- S_i^+ ) \rangle_{\rm inelastic}
= S + (2S-1) n - 2n^2,
\end{align}
where Eq. (\ref{eqn:t=0}) was used.
We can obtain the same result on the B sublattice
and summarize the total moment sum rule in Table \ref{table:S}.
\cite{Huberman-2005}

\begin{table}[t]
\caption{
Total moment sum rule obtained by the conventional spin-wave theory.
\cite{Huberman-2005}
Component of the dynamical spin correlation function $S^{\alpha\alpha}(\bq,\omega)$
and the integrated intensity $(1/N) \sum_\bq \int d\omega S^{\alpha\alpha}(\bq,\omega)$ are shown.
Since the $z$-axis is taken along the ordered moment, the $zz$ component is for longitudinal spin fluctuation,
while the $xx$ and $yy$ components are for transverse one.
The ordered moment at each site is expressed as $\langle S^z \rangle = S - n$ with
$n = \langle a^\dagger a \rangle = \langle b^\dagger b \rangle$.
We can obtain $S(S+1)$ after adding all components of the intensity.
Notice that the inelastic component $S^{zz}(\bq,\omega)_{\rm inelastic}$ represents two-magnon process.
}
\begin{tabular}{ccccc}
\hline
Component & Integrated intensity \\
\hline
$S^{zz}(\bq,\omega)_{\rm elastic}$ & $S^2 - 2S n + n^2$ \\
$S^{zz}(\bq,\omega)_{\rm inelastic}$ & $n + n^2$ \\
\hline
$[S^{xx}(\bq,\omega)+S^{yy}(\bq,\omega)]_{\rm inelastic}$ & $S+(2S-1)n-2n^2$ \\
\hline
$\sum_{\alpha=x,y,z} [ S^{\alpha\alpha}(\bq,\omega) ]_{\rm total}$ & $S(S+1)$ \\
\hline
\end{tabular}
\label{table:S}
\end{table}

\section{Extended Spin-Wave Theory}

The extended spin-wave theory, or the generalized Holstein-Primakoff theory,
\cite{Papanicolaou-1984,Onufrieva-1985,Joshi-1999}
is useful to describe excitations not only in quantum spin systems
but also in $f$ electron systems in multipole ordered states.
\cite{Shiina-2003,Shiina-2004}
In the formulation, plural bosons are introduced for each multiplet.
This enables us to describe magnetic excitations in quantum spin exhibiting a quantum phase transition,
where both the T- and L-modes are taken into account.
In the following, we introduce the formulation according to Ref. \ref{ref:Shiina-2003}.

\subsection{Formulation}

On the basis of the extended spin-wave theory,
we can simply estimate contributions form the elastic and inelastic components.
First, we briefly introduce the extended spin-wave theory.
As in the conventional spin-wave theory, we consider the Heisenberg Hamiltonian given by Eq. (\ref{eqn:H}).
On the A sublattice, we introduce a local state $|m\rangle_i$ as
\begin{align}
S_i^z |m\rangle_i = m |m\rangle_i.
\end{align}
We assume that $\hbar=1$ throughout this paper.
In the N\'{e}el ordered state, $|m=S\rangle_i$ is the local ground state, while $|m<S\rangle_i$ are excited states.
The spin operator on the A sublattice is then expressed as
\cite{Papanicolaou-1984,Onufrieva-1985,Joshi-1999,Shiina-2003}
\begin{align}
\bS_i = \sum_{m,n} | m \rangle_i ~_i\langle m | \bS_i | n \rangle_i ~_i\langle n |
= \sum_{m,n} ~_i\langle m | \bS_i | n \rangle_i a_{im}^\dagger a_{in}.
\label{eqn:S}
\end{align}
Here, $a_{im}^\dagger$ and $a_{in}$ are bosonic operators that create and annihilate the $|m\rangle_i$ and $| n\rangle_i$ states, respectively.
The bosons are subjected by the following local constraint:
\cite{Sachdev-1990,Shiina-2003}
\begin{align}
\sum_{m} a_{im}^\dagger a_{im} = 1.
\end{align}
We can see that Eq. (\ref{eqn:S}) contains $a_{iS}^\dagger a_{iS}$ term.
With the constraint, this term is expressed by bosons for the excited states as
\begin{align}
&a_{iS}^\dagger a_{iS} = M - \sum_{m<S} a_{im}^\dagger a_{im}.
\label{eqn:aS-0}
\end{align}
Here, we introduced $M$ as an expansion parameter for the theory, which we can finally put $M=1$ in the formulation.
Equation (\ref{eqn:S}) also contains $a_{in}^\dagger a_{iS}$ and $a_{iS}^\dagger a_{in}$ $(n < S)$ terms.
For these, it is convenient to introduce the following generalized Holstein-Primakoff method:
\cite{Papanicolaou-1984,Onufrieva-1985,Joshi-1999,Shiina-2003}
\begin{equation}
\begin{aligned}
&a_{in}^\dagger a_{iS} \rightarrow a_{in}^\dagger \left( M - \sum_{m<S} a_{im}^\dagger a_{im} \right)^{\frac{1}{2}}, \cr
&a_{iS}^\dagger a_{in} \rightarrow \left( M - \sum_{m<S} a_{im}^\dagger a_{im} \right)^{\frac{1}{2}} a_{in}.
\end{aligned}
\label{eqn:aS}
\end{equation}
When we substitute Eqs. (\ref{eqn:aS-0}) and (\ref{eqn:aS}) into Eq. (\ref{eqn:S}),
the spin operator is written by bosons for the excited states.
We next expand the square root in Eq. (\ref{eqn:aS}) and substitute the spin operator into the spin Hamiltonian.
The spin Hamiltonian is then expressed in powers of $M^{-1}$ as
\cite{Shiina-2003}
\begin{align}
\H = M^2 \sum_{n=0}^\infty M^{-\frac{n}{2}} \H_n.
\end{align}
Here, $\H_n$ represents $n$-boson Hamiltonian.
$\H_0$ is a c-number term and represents the energy of the mean-field ground state.
$\H_1$ is the first-order term of bosons for the excited states, however, it vanishes with the proper mean-field ground state.
$\H_2$ is the second-order (harmonic) term.
Since bosons are dilute at low temperatures, we neglect higher-order terms of bosons for the harmonic theory and put $M=1$.

The spin-wave excitation is described by the $a_{i,S-1}$ operator for $|m=S-1\rangle_i$.
Substituting Eq. (\ref{eqn:aS}) into Eq. (\ref{eqn:S}) and retaining up to the quadratic order of the $a_{i,S-1}$ boson,
we can express the spin operators as
\begin{align}
&S_i^x = \sqrt{S/2} ( a_{i,S-1} + a_{i,S-1}^\dagger ), \cr
&S_i^y = - i \sqrt{S/2} ( a_{i,S-1} - a_{i,S-1}^\dagger ),
\label{eqn:SA} \\
&S_i^z = S - a_{i,S-1}^\dagger a_{i,S-1}.
\nonumber
\end{align}
These results are same as those obtained by the linear spin-wave theory.

On the B sublattice, we introduce $b_{im}$ bosons for the local excited state.
$|m=-S\rangle_i$ is the local ground state, while $|m>-S\rangle_i$ are excited states.
In the same way as on the A sublattice, the spin operators on the B sublattice are expressed as
\begin{align}
&S_i^x = \sqrt{S/2} ( b_{i,-S+1} + b_{i,-S+1}^\dagger ), \cr
&S_i^y = i \sqrt{S/2} ( b_{i,-S+1} - b_{i,-S+1}^\dagger ),
\label{eqn:SB} \\
&S_i^z = - S + b_{i,-S+1}^\dagger b_{i,-S+1}.
\nonumber
\end{align}
We notice the following correspondences between the extended and conventional spin-wave theories:
\begin{align}
a_{i,S-1} \leftrightarrow a_i,~~~~~~
b_{i,-S+1} \leftrightarrow b_i.
\end{align}
Substituting Eqs. (\ref{eqn:SA}) and (\ref{eqn:SB}) into Eq. (\ref{eqn:H}),
we can reproduce the Hamiltonian in Eq. (\ref{eqn:H2}) for the linear spin-wave theory.

\subsection{Application to total moment sum rule}

\subsubsection{Ordered moment}

To understand how to apply the extended spin-wave theory to the total moment sum rule, we focus on an $S=1$ spin case.
Since the result is the same on the A and B sublattices, we focus on the A sublattice.
The local ground state is $|1\rangle$, whereas $|0\rangle$ and $|-1\rangle$ are excited states.
The linear spin-wave theory is described by $a_0$ boson for the $|0\rangle$ state.
Within the basal $|1\rangle$, $|0\rangle$, and $|-1\rangle$ states, $S^z$ is expressed in the following matrix form:
\begin{align}
S^z =
\begin{pmatrix}
1 & 0 & 0 \cr
0 & 0 & 0 \cr
0 & 0 & -1
\end{pmatrix}
\end{align}
In the extended spin-wave theory, $S^z$ is then expressed as
\begin{align}
S^z = a_{1}^\dagger a_{1} - a_{-1}^\dagger a_{-1}.
\end{align}
Using the local constraint
\begin{align}
a_{1}^\dagger a_{1} + a_{0}^\dagger a_{0} + a_{-1}^\dagger a_{-1} = 1,
\label{eqn:constraint-1}
\end{align}
we eliminate the $a_{1}^\dagger a_{1}$ term for the local ground state and obtain
\begin{align}
S^z = 1 - a_{0}^\dagger a_{0} - 2 a_{-1}^\dagger a_{-1}.
\label{eqn:Sz-extend}
\end{align}
Thus, $S^z$ is expressed by the bosons for the excited states.
In the matrix representation, Eq. (\ref{eqn:Sz-extend}) is equivalent to
\begin{align}
S^z =
\begin{pmatrix}
1 & 0 & 0 \cr
0 & 1 & 0 \cr
0 & 0 & 1
\end{pmatrix}
-
\begin{pmatrix}
0 & 0 & 0 \cr
0 & 1 & 0 \cr
0 & 0 & 2
\end{pmatrix}
= {\bf{1}} - \Delta S^z,
\label{eqn:Sz-matrix}
\end{align}
where
\begin{align}
{\bf{1}} =
\begin{pmatrix}
1 & 0 & 0 \cr
0 & 1 & 0 \cr
0 & 0 & 1
\end{pmatrix},~~~
\Delta S^z =
\begin{pmatrix}
0 & 0 & 0 \cr
0 & 1 & 0 \cr
0 & 0 & 2
\end{pmatrix}
=
a_0^\dagger a_0 + 2 a_{-1}^\dagger a_{-1}.
\label{eqn:sz-matrix}
\end{align}
The expectation value of the ordered moment is calculated as
\begin{align}
\langle S^z \rangle
= 1 - \langle a_{0}^\dagger a_{0} \rangle
= 1 - n.
\end{align}
Here, we used $\langle a_{-1}^\dagger a_{-1} \rangle = 0$
and put $n = \langle a_{0}^\dagger a_{0} \rangle$.
It does not depend on the spin site.

\subsubsection{$z$ component}

Next, we discuss the integrated dynamical spin correlation function, as shown in Eq. (\ref{eqn:Szz}).
Using Eq. (\ref{eqn:Sz-extend}), we can express the operator $S^z S^z(t)$ as
\begin{align}
S^z S^z(t) &= \left[ 1 - a_{0}^\dagger a_{0} - 2 a_{-1}^\dagger a_{-1} \right]
                          \left[ 1 - a_{0}^\dagger(t) a_{0}(t)- 2 a_{-1}^\dagger(t) a_{-1}(t) \right] \cr
&\rightarrow \left[ 1 - a_{0}^\dagger a_{0} \right] \left[ 1 - a_{0}^\dagger(t) a_{0}(t) \right] \cr
&= 1 - a_{0}^\dagger a_{0} - a_{0}^\dagger(t) a_{0}(t) + a_{0}^\dagger a_{0} a_{0}^\dagger(t) a_{0}(t).
\label{eqn:Szz-t}
\end{align}
Here, we dropped the site index $i$, since the expectation value of the operator is homogeneous and does not depend on the site.
We also dropped the $a_{-1}^\dagger a_{-1}$ terms, since its expectation value vanishes.
As in Eq. (\ref{eqn:Szz2}), the expectation value of $S^z S^z$ is separated into the elastic and inelastic components as
\begin{align}
\langle S^z S^z \rangle
&= 1 - 2 \langle a_{0}^\dagger a_{0} \rangle_{\rm elastic} + \langle a_{0}^\dagger a_{0} a_{0}^\dagger a_{0} \rangle \cr
&= 1 - 2 n + \langle a_{0}^\dagger a_{0} \rangle \langle a_{0}^\dagger a_{0} \rangle_{\rm elastic}
+ \langle a_{0}^\dagger a_{0} \rangle \langle a_{0} a_{0}^\dagger \rangle_{\rm 2-magnon} \cr
&= 1 - 2 n + n^2 + n(1+n).
\label{eqn:Szz-extend}
\end{align}
Here, we put $t=0$ in Eq. (\ref{eqn:Szz-t}) assuming after the integral over $\omega$,
and used Eqs. (\ref{eqn:t=0}) and (\ref{eqn:contraction}) for the expectation value of
$\langle a_{0}^\dagger a_{0} a_{0}^\dagger a_{0} \rangle$.
The first three terms, $1-2n+n^2$, are elastic component,
whereas the last term, $n(1+n)$, is inelastic component (two-magnon process), as discussed below Eq. (\ref{eqn:a4}).

The value of $n$ can be expected to be small.
When we neglect the $O(n^2)$ order terms, we can obtain the result in Eq. (\ref{eqn:Szz-extend}) in a simple way.
Up to the $O(n)$ order, the $\langle a_{0}^\dagger a_{0} a_{0}^\dagger a_{0} \rangle$ term in Eq. (\ref{eqn:Szz-extend}) can be treated as
\begin{align}
\langle a_{0}^\dagger a_{0} a_{0}^\dagger a_{0} \rangle
&= \langle a_{0}^\dagger a_{0} \rangle \langle a_{0}^\dagger a_{0} \rangle_{\rm elastic}
  + \langle a_{0}^\dagger a_{0} \rangle \langle a_{0} a_{0}^\dagger \rangle_{\rm 2-magnon} \cr
&\rightarrow \langle a_{0}^\dagger a_{0} \rangle \langle a_{0} a_{0}^\dagger \rangle_{\rm 2-magnon} \cr
&\rightarrow \langle a_{0}^\dagger a_{0} \rangle_{\rm 2-magnon}.
\end{align}
This corresponds to the following replacement:
\begin{align}
a_{0}^\dagger a_{0} a_{0}^\dagger a_{0} \rightarrow a_{0}^\dagger a_{0},
\label{eqn:treatment-1}
\end{align}
where the $O(n^2)$ order term was neglected.
In a matrix form, Eq. (\ref{eqn:treatment-1}) is equivalent to
\begin{align}
|0\rangle \langle 0 | 0\rangle \langle 0 | \rightarrow |0\rangle \langle 0 |.
\label{eqn:treatment-2}
\end{align}
To see this point, let us calculate $S^z S^z$ with Eq. (\ref{eqn:Sz-matrix}) as
\begin{align}
S^z S^z
= \left( {\bf{1}} - \Delta S^z \right)^2
= {\bf{1}} - 2 \Delta S^z  + \left(\Delta S^z\right)^2.
\label{eqn:Szz-matrix}
\end{align}
Using the matrix form in Eq. (\ref{eqn:sz-matrix}), we can calculate
\begin{align}
(\Delta S^z)^2=
\begin{pmatrix}
0 & 0 & 0 \cr
0 & 1 & 0 \cr
0 & 0 & 4
\end{pmatrix}
= a_0^\dagger a_0 + 4 a_{-1}^\dagger a_{-1}
\rightarrow a_0^\dagger a_0,
\end{align}
where we dropped the $a_{-1}^\dagger a_{-1}$ term.
When we take square of $\Delta S^z$ in the matrix form, the relation in Eq. (\ref{eqn:treatment-2}) was used implicitly.
\cite{note:sz}
In the representation with bosons, this is expressed as
\begin{align}
(\Delta S^z)^2=&\left( a_0^\dagger a_0 + 2 a_{-1}^\dagger a_{-1} \right)^2 \cr
&= a_0^\dagger a_0 a_0^\dagger a_0 + 4 a_{-1}^\dagger a_{-1} a_{-1}^\dagger a_{-1} + 4 a_0^\dagger a_0 a_{-1}^\dagger a_{-1} \cr
&\rightarrow a_0^\dagger a_0,
\end{align}
where we dropped the $a_{-1}^\dagger a_{-1}$ term and used Eq. (\ref{eqn:treatment-1}).
Therefore, the matrix representation is equivalent to that with bosons up to the $O(n)$ order.

Up the $O(n)$ order, the expectation value is calculated as
\begin{align}
\langle S^z S^z \rangle
&= \langle {\bf{1}} - 2 \Delta S^z  + \left(\Delta S^z\right)^2 \rangle \cr
&\rightarrow 1
- 2\langle
\begin{pmatrix}
0 & 0 & 0 \cr
0 & 1 & 0 \cr
0 & 0 & 2
\end{pmatrix}\rangle
+ \langle
\begin{pmatrix}
0 & 0 & 0 \cr
0 & 1 & 0 \cr
0 & 0 & 4
\end{pmatrix}\rangle \cr
&= 1_{\rm elastic} - 2 \langle a_0^\dagger a_0 \rangle_{\rm elastic} + \langle a_0^\dagger a_0 \rangle_{\rm 2-magnon} \cr
&= 1 - 2 n + n.
\end{align}
Thus, up to the $O(n)$ order, we can simply calculate the expectation value
by using the matrices of operators and their bosonic expressions within the extended spin-wave theory.

\subsubsection{$x$ and $y$ components}

We calculate the integrated correlation function of the $x$ and $y$ components
up to the $O(n)$ order within the matrix form of the spin operators.
For $S=1$, $S^x S^x+S^y S^y$ is expressed as
\begin{align}
S^x S^x + S^y S^y &=
\begin{pmatrix}
1 & 0 & 0 \cr
0 & 2 & 0 \cr
0 & 0 & 1
\end{pmatrix} \cr
&= a_1^\dagger a_1 + a_0^\dagger a_0 + a_0^\dagger a_0 + a_{-1}^\dagger a_{-1}.
\end{align}
The expectation value is calculated as
\begin{align}
&\langle S^x S^x + S^y S^y \rangle \cr
&=\langle a_1^\dagger a_1 \rangle
+ \langle a_0^\dagger a_0 \rangle + \langle a_0^\dagger a_0 \rangle + \langle a_{-1}^\dagger a_{-1} \rangle
\label{eqn:ref} \\
&=
\langle 1 - a_0^\dagger a_0 - a_{-1}^\dagger a_{-1} \rangle
+ \langle a_0^\dagger a_0 \rangle + \langle a_0^\dagger a_0 \rangle + \langle a_{-1}^\dagger a_{-1} \rangle \cr
&= 1 + n.
\label{eqn:Sxy}
\end{align}
Here, we used the local constraint in Eq. (\ref{eqn:constraint-1}) and eliminated the $a_1^\dagger a_1$ term for the local ground state.
The first term in Eq. (\ref{eqn:ref}) originates from $\langle a_1^\dagger a_0 a_0^\dagger a_1 \rangle$
and represents a one-magnon process.
The second term is from $\langle a_0^\dagger a_1 a_1^\dagger a_0 \rangle$ and is a one-magnon process.
The third term is from $\langle a_0^\dagger a_{-1} a_{-1}^\dagger a_0 \rangle$ and is a multi-magnon process.
In the conventional spin-wave theory, the $a_0$ boson describes one-magnon excitations,
whereas the $a_{-1}$ boson is for two-magnon excitations.
Therefore, the $\langle a_0^\dagger a_{-1} a_{-1}^\dagger a_0 \rangle$ term
can be understood as a multi-magnon (three-magnon) process in terms of the conventional spin-wave theory.
The last term in Eq. (\ref{eqn:ref}) is from $\langle a_{-1}^\dagger a_{0} a_{0}^\dagger a_{-1} \rangle$
and is a multi-magnon process.
However, this term vanishes because $\langle a_{-1}^\dagger a_{-1} \rangle=0$.

Thus, the matrix form of the spin operators and their bosonic representations in the extended spin-wave theory
is useful to resolve the total moment sum rule into elastic and inelastic components.
We can also resolve the inelastic component into one-magnon and multi-magnon processes.
Here, we demonstrated this point with the $S=1$ case.
Notice that this formulation can be extended to a general value of $S$.
The result reproduces Table \ref{table:S} up to the $O(n)$ order, and we summarize it in Table \ref{table:S=S}.

\begin{table}[t]
\caption{
Total moment sum rule obtained by the extended spin-wave theory up to the $O(n)$ order for spin $S$ systems.
Components of the dynamical spin correlation function and the integrated intensities are shown.
Since the $z$-axis taken along the ordered moment, the $zz$ component is for longitudinal spin fluctuation,
while the $xx$ and $yy$ components are for transverse one.
The moment per one site is expressed as $\langle S^z \rangle = S - n$ with
$n= \langle a_{S-1}^\dagger a_{S-1} \rangle = \langle b_{-S+1}^\dagger b_{-S+1} \rangle$.
We can obtain $S(S+1)$ after adding all components of the intensity.
The result reproduces Table \ref{table:S} for the conventional spin-wave theory.
}
\begin{tabular}{ccccc}
\hline
Component & Integrated intensity \\
\hline
$S^{zz}(\bq,\omega)_{\rm elastic}$ & $S^2 - 2Sn$ \\
$S^{zz}(\bq,\omega)_{\rm two-magnon}$ & $n$ \\
\hline
$[S^{xx}(\bq,\omega)+S^{yy}(\bq,\omega)]_{\rm 1-magnon}$ & $S$ \\
$[S^{xx}(\bq,\omega)+S^{yy}(\bq,\omega)]_{\rm multi-magnon}$ & $(2S-1)n$ \\
\hline
$\sum_{\alpha=x,y,z}[ S^{\alpha\alpha}(\bq,\omega) ]_{\rm total}$ & $S(S+1)$ \\
\hline
\end{tabular}
\label{table:S=S}
\end{table}

\section{$S=1$ Systems With Single-Ion Anisotropy}

We first study a simple system with a quantum phase transition.
Let us begin with the following spin Hamiltonian on the simple cubic lattice:
\begin{align}
\H = D\sum_i (S_i^z)^2 + J \sum_{\langle i,j\rangle} \bS_i \cdot \bS_j.
\label{eqn:HD}
\end{align}
Here, $D(>0)$ represents the single-ion anisotropy of easy-plane type.
For $S=1$, the local energy states are split into singlet ($S^z=0$) and doublet ($S^z=\pm 1$) states.
There is a quantum critical point which separates disordered and ordered phases at zero temperature.
For large (small) $D/J$, the disordered (ordered) phase is stabilized.
We study the total moment sum rule based on the extended spin-wave theory.

\subsection{Mean-field solution}

Under the easy-plane anisotropy, the ordered moment lies in the $xy$-plane.
We take the $x$-axis along the ordered moment.
The mean-field Hamiltonian is then given by
\begin{align}
\H_{\rm MF} = D(S^z)^2 - 6J \langle S^x \rangle S^x.
\label{eqn:H-mf}
\end{align}
The factor $6J$ is from the simple cubic lattice.
In the presence of a molecular field in the $x$ direction, the twofold degeneracy of the doublet is lifted
and the $|0\rangle$ and $|\pm 1\rangle$ states are hybridized.
The local ground and excited states at the $i$th site are expressed in the following form:
\cite{Matsumoto-2007,Matsumoto-2014}
\begin{align}
&|{\rm G}\rangle_i = u |0\rangle_i + v_i \frac{1}{\sqrt{2}}( |1\rangle_i + |-1\rangle_i ), \cr
&|{\rm T}\rangle_i = \frac{1}{\sqrt{2}}(- |1\rangle_i + |-1\rangle_i ),
\label{eqn:gs} \\
&|{\rm L}\rangle_i = -v_i |0\rangle_i + u \frac{1}{\sqrt{2}}( |1\rangle_i + |-1\rangle_i ).
\nonumber
\end{align}
Here, $|{\rm G}\rangle_i$ represents the ground state,
while $|{\rm T}\rangle_i$ and $|{\rm L}\rangle_i$ are excited states.
The former and the latter states are for transverse and longitudinal modes (L- and T-modes), respectively.
We will discuss this point later.
In Eq. (\ref{eqn:gs}), we introduced
\begin{align}
v_i=e^{i\bQ\cdot\br_i}v,
\end{align}
where $u$ and $v$ are real coefficients satisfying $u^2+v^2=1$.
$\bQ=(\pi,\pi,\pi)$ is the AF wave vector and $\br_i$ represents the position of the $i$th site.
Thus, $e^{i\bQ\cdot\br_i}=\pm 1$ on the A and B sublattices, respectively.
The expectation value of the spin operator $\bS_i$ is
\begin{align}
~_i\langle {\rm G}| \bS_i |{\rm G}\rangle_i = 2uv_i \be_x,
\end{align}
with $\be_x$ as a unit vector along the $x$ direction.
The expectation value in Eq. (\ref{eqn:H-mf}) is then given by
\begin{align}
\langle S_i^x \rangle=2uv_i.
\end{align}

The mean-field energy per one site is given by
\begin{align}
E_{\rm MF} &= \langle {\rm G}| [ D(S^z)^2 - 3J (2uv) S^x ] |{\rm G}\rangle \cr
&= ( D - J_{\rm eff} ) v^2 + J_{\rm eff} v^4,
\label{eqn:H-mf-av}
\end{align}
where 
\begin{align}
J_{\rm eff}=12J.
\end{align}
The coefficients $u$ and $v$ are determined so as to minimize the mean-field energy.
For $J_{\rm eff}/D\ge 1$, they are determined as
\begin{align}
u = \sqrt{ \frac{1}{2}\left( 1 + \frac{D}{J_{\rm eff}} \right)},~~~~~~
v = \sqrt{ \frac{1}{2}\left( 1 - \frac{D}{J_{\rm eff}} \right)}.
\label{eqn:uv-D}
\end{align}
For $J_{\rm eff}/D\le 1$, $u=1$ and $v=0$.
Therefore, $J_{\rm eff}=D$ represents a quantum critical point at which the disorder and ordered phases are separated.

\subsection{Extended spin-wave theory}

\begin{figure}
\begin{center}
\includegraphics[width=4cm]{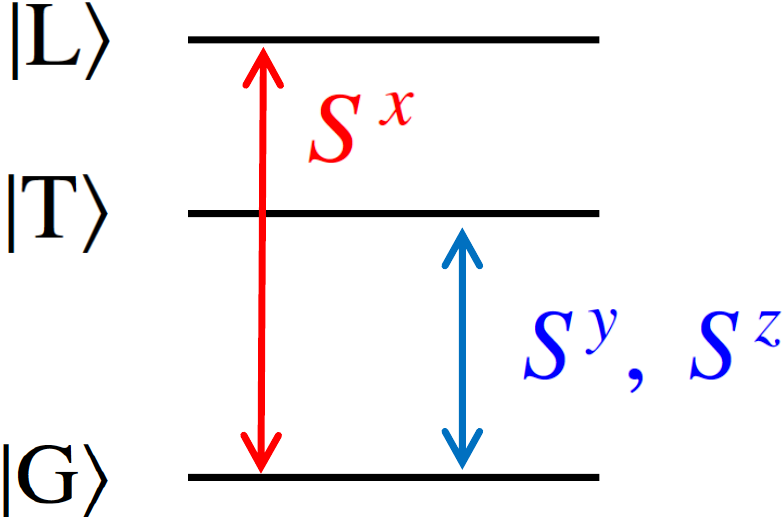}
\end{center}
\caption{
(Color online)
Schematic of energy levels and matrix elements of spin operators.
Here, the $x$-axis is taken along the ordered moment.
$S^x$ has a finite matrix element between the ground ($|{\rm G}\rangle$) and the $|{\rm L}\rangle$ excited states.
$S^y$ and $S^z$ have a finite matrix element between the ground and the $|{\rm T}\rangle$ excited states.
Since $S^x$ and $(S^y,S^z)$ are the longitudinal and transverse components
and lead to longitudinal and transverse fluctuations of the ordered moment,
the $|{\rm L}\rangle$ and $|{\rm T}\rangle$ states are for the L- and T-modes, respectively.
}
\label{fig:L-mode}
\end{figure}

Next, we represent the spin operators on the basis of the
$|{\rm G}\rangle_i$, $|{\rm T}\rangle_i$, and $|{\rm L}\rangle_i$ states.
The $S_i^x$ spin operator is expressed in the following matrix form:
\cite{Papanicolaou-1984,Matsumoto-2007}
\begin{align}
&S_i^x
=
\begin{pmatrix}
2uv_i & 0 & u^2-v^2 \cr
0 & 0 & 0 \cr
u^2-v^2 & 0 & -2uv_i
\end{pmatrix} \cr
&=
\begin{pmatrix}
2uv_i & 0 & 0 \cr
0 & 2uv_i & 0 \cr
0 & 0 & 2uv_i
\end{pmatrix}
+
\begin{pmatrix}
0 & 0 & u^2-v^2 \cr
0 & -2uv_i & 0 \cr
u^2-v^2 & 0 & -4uv_i
\end{pmatrix} \cr
&= 2uv_i + (u^2-v^2) ( a_{iG}^\dagger a_{iL} + a_{iL}^\dagger a_{iG} )
- 2uv_i a_{iT}^\dagger a_{iT} - 4uv_i a_{iL}^\dagger a_{iL} \cr
&\rightarrow 2uv_i + (u^2-v^2) ( a_{iL} + a_{iL}^\dagger )
- 2uv_i a_{iT}^\dagger a_{iT} - 4uv_i a_{iL}^\dagger a_{iL}.
\label{eqn:SxD}
\end{align}
Here, we introduced $a_{iG}$, $a_{iT}$, and $a_{iL}$ bosons
for the $|{\rm G}\rangle_i$, $|{\rm T}\rangle_i$, and $|{\rm L}\rangle_i$ states, respectively.
We used $a_{iG}\rightarrow ( 1 - a_{iT}^\dagger a_{iT} - a_{iL}^\dagger a_{iL} )^{\frac{1}{2}}$
and retained up to the quadratic order of $a_T$ and $a_L$.
We can see in Eq. (\ref{eqn:SxD}) that $S_i^x$ has a finite matrix elements
between the $|{\rm G}\rangle_i$ and $|{\rm L}\rangle_i$ states (see Fig. \ref{fig:L-mode}).
This represents that the $|{\rm L}\rangle_i$ state has a spin fluctuation
parallel to the ordered moment along the $x$ direction.
Therefore, the $b_{iL}$ boson describes the longitudinal excitation mode.

The $S_i^y$ and $S_i^z$ operators are expressed as
\begin{equation}
\begin{aligned}
S_i^y
&=
\begin{pmatrix}
0 & -iu & 0 \cr
iu & 0 & -iv_i \cr
0 & iv_i & 0
\end{pmatrix} \cr
&= -iu ( a_{iG}^\dagger a_{iT} - a_{iT}^\dagger a_{iG} ) - iv_i ( a_{iT}^\dagger a_{iL} - a_{iL}^\dagger a_{iT} ) \cr
&\rightarrow -iu ( a_{iT} - a_{iT}^\dagger ) - iv_i ( a_{iT}^\dagger a_{iL} - a_{iL}^\dagger a_{iT} ), \cr
S_i^z
&=
\begin{pmatrix}
0 & -v_i & 0 \cr
-v_i & 0 & -u \cr
0 & -u & 0
\end{pmatrix} \cr
&= - v_i ( a_{iG}^\dagger a_{iT} + a_{iT}^\dagger a_{iG} ) - u ( a_{iT}^\dagger a_{iL} + a_{iL}^\dagger a_{iT} ) \cr
&\rightarrow - v_i ( a_{iT} + a_{iT}^\dagger ) - u ( a_{iT}^\dagger a_{iL} + a_{iL}^\dagger a_{iT} ).
\end{aligned}
\label{eqn:SyzD}
\end{equation}
We can see that both $S_i^y$ and $S_i^z$ have a finite matrix element
between the $|{\rm G}\rangle_i$ and $|{\rm T}\rangle_i$ states (see Fig. \ref{fig:L-mode}).
Therefore, the $b_{iT}$ boson describes the transverse excitation mode.
For the $D$ term in Eq. (\ref{eqn:HD}), $(S_i^z)^2$ is expressed as
\begin{align}
&(S_i^z)^2 =
\begin{pmatrix}
v^2 & 0 & uv_i \cr
0 & v^2+u^2 & 0 \cr
uv_i & 0 & u^2
\end{pmatrix} \cr
&~~~
= v^2 {\bm{1}} +
\begin{pmatrix}
0 & 0 & uv_i \cr
0 & u^2 & 0 \cr
uv_i & 0 & u^2 - v^2
\end{pmatrix} \cr
&~~~
= v^2 + u^2 a_{iT}^\dagger a_{iT} + (u^2-v^2) a_{iL}^\dagger a_{iL} + uv_i ( a_{iL} + a_{iL}^\dagger ).
\label{eqn:Sz2}
\end{align}

Now, the spin operators are expressed with the bosons for the excited states.
Substituting Eqs. (\ref{eqn:SxD}), (\ref{eqn:SyzD}), and (\ref{eqn:Sz2}) into Eq. (\ref{eqn:HD}), we obtain
\cite{Matsumoto-2007,Matsumoto-2014}
\begin{align}
\H = \sum_\bk &\sum_{m=T,L} \left[ \epsilon_{\bk m} a_{\bk m}^\dagger a_{\bk m}
+ \frac{1}{2} \Delta_{\bk m} ( a_{\bk m} a_{\bk m} + a_{\bk m}^\dagger a_{\bk m}^\dagger ) \right],
\label{eqn:H-D}
\end{align}
where
\begin{align}
&\epsilon_{\bk T} = u^2 D + 2(uv)^2 J_{\rm eff} + (u^2-v^2)\gamma_\bk, \cr
&\epsilon_{\bk L} = (u^2-v^2)D + 4(uv)^2 J_{\rm eff} + (u^2-v^2)^2\gamma_\bk, \cr
&\Delta_{\bk T} = - \gamma_\bk,
\label{eqn:ed-D} \\
&\Delta_{\bk L} = (u^2-v^2)^2 \gamma_\bk, \cr
&\gamma_\bk = 2J ( \cos{k_x} + \cos{k_y} + \cos{k_z} ).
\nonumber
\end{align}
In Eq. (\ref{eqn:H-D}), we retained up to the quadratic order of the bosons.
Note that the first order term of the Bose operator vanishes by Eq. (\ref{eqn:uv-D}).
$a_{\bk m}$ is the Fourier transformed operator of $a_{im}$ ($m=T,L$).
Introducing the following Bogoliubov transformation,
\begin{equation}
\begin{aligned}
&a_{\bk m} = u_{\bk m} \alpha_{\bk m} + v_{\bk m} \alpha_{-\bk m}^\dagger, \cr
&u_{\bk m} = \sqrt{ \frac{1}{2} \left( \frac{\epsilon_{\bk m}}{E_{\bk m}} + 1 \right)}, \cr
&v_{\bk m} = \sqrt{ \frac{1}{2} \left( \frac{\epsilon_{\bk m}}{E_{\bk m}} - 1 \right)} \frac{\Delta_{\bk m}}{|\Delta_{\bk m}|}, \cr
&E_{\bk m} = \sqrt{ \epsilon_{\bk m}^2 - \Delta_{\bk m}^2},
\end{aligned}
\label{eqn:uv-Bogo-D}
\end{equation}
we can diagonalize the Hamiltonian in Eq. (\ref{eqn:H-D}) as
\begin{align}
\H = \sum_\bk \sum_{m=T,L} E_{\bk m} \alpha_{\bk m}^\dagger \alpha_{\bk m}.
\end{align}
Here, we dropped a constant term.
As in Eq. (\ref{eqn:n}), the expectation value is calculated as
\begin{align}
n_m = \langle a_{im}^\dagger a_{im} \rangle
= \frac{1}{N} \sum_\bk v_{\bk m}^2.~~~(m=T,L)
\label{eqn:n-TL}
\end{align}

\begin{figure}
\begin{center}
\includegraphics[width=5cm]{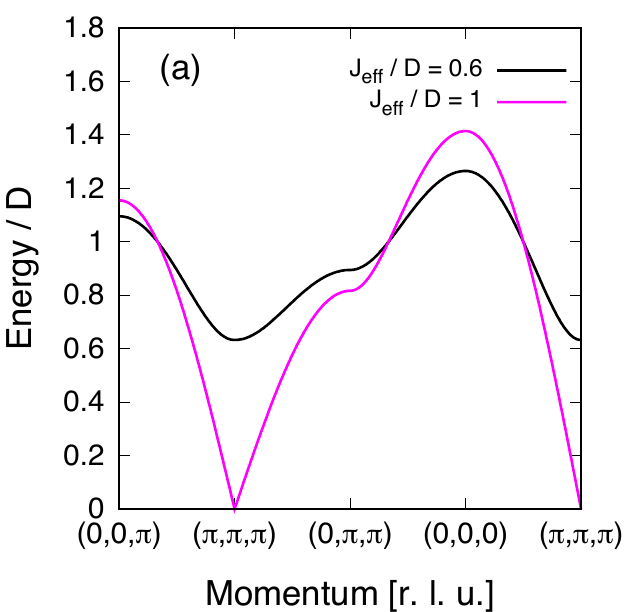}
\includegraphics[width=5cm]{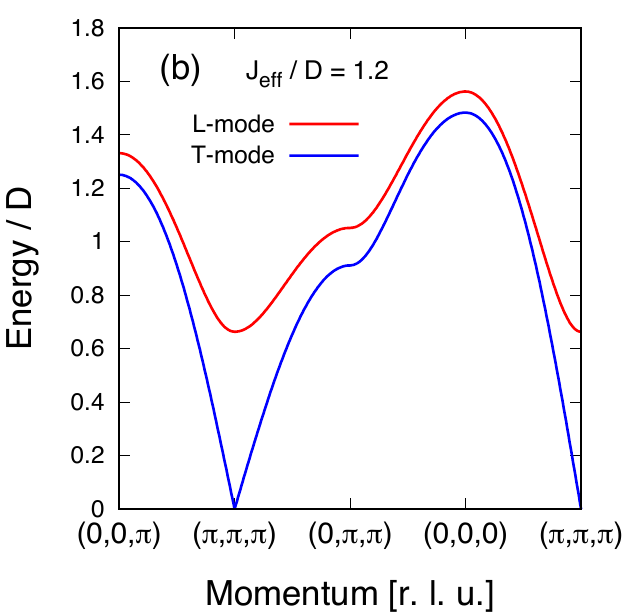}
\end{center}
\caption{
(Color online)
Magnon dispersion relation in S=1 systems on a simple cubic lattice with a single-ion anisotropy of easy-plane type.
(a) For $J_{\rm eff}/D \le 1$ (disordered phase).
(b) For $J_{\rm eff}/D > 1$ (ordered phase).
The excitation modes split into L- and T-modes.
}
\label{fig:S=1}
\end{figure}

For $J_{\rm eff}/D<1$, the disordered phase is stabilized.
The magnetic excitation is twofold degenerate with a finite excitation gap at the AF wave vector $\bQ=(\pi,\pi,\pi)$,
as shown in Fig. \ref{fig:S=1}(a).
At the quantum critical point, $J_{\rm eff}/D = 1$, the excitation becomes soft
showing a linear dispersion relation around $\bk=\bQ$.
For $J_{\rm eff}/D>1$, $v$ becomes finite and the ordered phase is stabilized,
where the twofold degenerate modes split into the T- and L-modes.
The T-mode is gapless, while the L-mode has a finite gap, as shown in Fig. \ref{fig:S=1}(b).
In the $J_{\rm eff}/D \rightarrow \infty$ limit, the L-mode moves to a high-energy region and becomes flat (dispersionless).
In the vicinity of the quantum critical point in the ordered state, on the other hand,
the L-mode is located in a low-energy region with strong intensity for inelastic neutron scattering.
\cite{Matsumoto-2007,Matsumoto-2014}
This appears as a substantial weight of the integrated correlation function of the L-mode in one-magnon process.

\subsection{Total moment sum rule}

\subsubsection{Ordered moment}

From Eq. (\ref{eqn:SxD}), we can calculate the expectation value of the staggered ordered moment per site as
\begin{align}
|\langle S_i^x \rangle| = 2uv ( 1 - n_\rT - 2n_\rL ),
\label{eqn:Sx-av}
\end{align}
where $n_\rT$ and $n_\rL$ are introduced in Eq. (\ref{eqn:n-TL}).
In Eq. (\ref{eqn:Sx-av}), $2uv$ is the mean-field value,
whereas the factor $( 1 - n_\rT - 2n_\rL)$ originates from the quantum correction.
Notice that the mean-field value takes $2uv=1$ ($2uv=0$) for
$J_{\rm eff}/D\rightarrow \infty$ ($J_{\rm eff}/D\rightarrow 1$).
The important point is that the moment already shrinks in the mean-field level
owing to the single-ion anisotropy of the easy-plane type.

\subsubsection{$x$ component}

From Eq. (\ref{eqn:SxD}), $(S_i^x)^2$ is expressed as
\begin{align}
(S_i^x)^2
&=
(2uv)^2 {\bm{1}}
+ 2(2uv_i)
\begin{pmatrix}
0 & 0 & u^2-v^2 \cr
0 & -2uv_i & 0 \cr
u^2-v^2 & 0 & -4uv_i
\end{pmatrix} \cr
&~~~
+
\begin{pmatrix}
0 & 0 & u^2-v^2 \cr
0 & -2uv_i & 0 \cr
u^2-v^2 & 0 & -4uv_i
\end{pmatrix}^2.
\end{align}
Here, the first two terms are elastic component,
while the last term is inelastic one and can be expressed as
\begin{align}
&
\begin{pmatrix}
0 & 0 & u^2-v^2 \cr
0 & -2uv_i & 0 \cr
u^2-v^2 & 0 & -4uv_i
\end{pmatrix}^2 \cr
&=
\begin{pmatrix}
(u^2-v^2)^2 & 0 & -4uv_i(u^2-v^2) \cr
0 & (2uv)^2 & 0 \cr
-4uv_i(u^2-v^2) & 0 & (u^2-v^2)^2 + (4uv)^2
\end{pmatrix} \cr
&\xrightarrow{{\rm DP}}
(u^2-v^2)^2 a_{iG}^\dagger a_{iG} + (2uv)^2a_{iT}^\dagger a_{iT} \cr
&~~~~~~
+ (u^2-v^2)^2 a_{iL}^\dagger a_{iL} + (4uv)^2 a_{iL}^\dagger a_{iL}.
\label{eqn:Sx-D2}
\end{align}
Here, ``DP" means the diagonal part and we only retained the diagonal part,
since the off diagonal part vanishes after taking the expectation value.
In Eq. (\ref{eqn:Sx-D2}), the first term is from $a_{iG}^\dagger a_{iL} a_{iL}^\dagger a_{iG}$ and is a one-magnon process.
The expectation value is calculated as
$\langle a_{iG}^\dagger a_{iG}\rangle = \langle 1 - a_{iT}^\dagger a_{iT} - a_{iL}^\dagger a_{iL} \rangle = 1 - n_\rT - n_\rL$.
The second term is from $a_{iT}^\dagger a_{iT} a_{iT}^\dagger a_{iT}$ and it a two-magnon process.
The third term is from $a_{iL}^\dagger a_{iG} a_{iG}^\dagger a_{iL}$ and is a one-magnon process.
The last term is from $a_{iL}^\dagger a_{iL} a_{il}^\dagger a_{iL}$ and is a two-magnon process.
Thus, the expectation value can be expressed as
\begin{align}
&\langle S_i^x S_i^x \rangle_{\rm elastic} = (2uv)^2 - 2(2uv)^2 ( n_\rT + 2n_\rL ), \cr
&\langle S_i^x S_i^x \rangle_{\rm 1-magnon} = (u^2-v^2)^2 ( 1 - n_\rT ) , \\
&\langle S_i^x S_i^x \rangle_{\rm 2-magnon} = (2uv)^2 n_\rT + (4uv)^2 n_\rL.
\nonumber
\end{align}

\subsubsection{$y$ component}

From Eq. (\ref{eqn:SyzD}), the $(S_i^y)^2$ is expressed as
\begin{align}
(S_i^y)^2
&=
\begin{pmatrix}
u^2 & 0 & -uv_i \cr
0 & u^2+v^2 & 0 \cr
-uv_i & 0 & v^2
\end{pmatrix} \cr
&\xrightarrow{{\rm DP}}
u^2 a_{iG}^\dagger a_{iG} + u^2 a_{iT}^\dagger a_{iT}
+ v^2 a_{iT}^\dagger a_{iT} + v^2 a_{iL}^\dagger a_{iL}.
\label{eqn:y-av}
\end{align}
The first term is from $a_{iG}^\dagger a_{iT} a_{iT}^\dagger a_{iG}$ and is a one-magnon process.
The second term is from $a_{iT}^\dagger a_{iG} a_{iG}^\dagger a_{iT}$ and is a one-magnon process.
The third term is from $a_{iT}^\dagger a_{iL} a_{iL}^\dagger a_{iT}$ and is a two-magnon process.
The last term is from $a_{iL}^\dagger a_{iT} a_{iT}^\dagger a_{iL}$ and is a two-magnon process.
Since $\langle S_i^y \rangle=0$, there is no elastic term.
Therefore, the expectation value are expressed as
\begin{equation}
\begin{aligned}
&\langle S_i^y S_i^y \rangle_{\rm 1-magnon} = u^2 ( 1 - n_\rL), \cr
&\langle S_i^y S_i^y \rangle_{\rm 2-magnon} = v^2 ( n_\rT + n_\rL ).
\end{aligned}
\end{equation}

\subsubsection{$z$ component}

From Eq. (\ref{eqn:Sz2}), $(S_i^z)^2$ is expressed as
\begin{align}
(S_i^z)^2 &=
\begin{pmatrix}
v^2 & 0 & uv_i \cr
0 & v^2+u^2 & 0 \cr
uv_i & 0 & u^2
\end{pmatrix} \cr
&\xrightarrow{{\rm DP}}
v^2 a_{iG}^\dagger a_{iG} + v^2 a_{iT}^\dagger a_{iT}
+ u^2 a_{iT}^\dagger a_{iT} + u^2 a_{iL}^\dagger a_{iL}.
\end{align}
This is essentially the same as Eq. (\ref{eqn:y-av}), and the expectation values are expressed as
\begin{equation}
\begin{aligned}
&\langle S_i^z S_i^z \rangle_{\rm 1-magnon} = v^2 ( 1 - n_\rL), \cr
&\langle S_i^z S_i^z \rangle_{\rm 2-magnon} = u^2 ( n_\rT + n_\rL ).
\end{aligned}
\end{equation}

In Table \ref{table:S=1-D}, we summarize the result.
We emphasize that the longitudinal spin fluctuation component $S^{xx}(\bq,\omega)$
has a finite value in the one-magnon process by the L-mode.
In the $J_{\rm eff}/D \rightarrow \infty$ limit, $u=v=1/\sqrt{2}$ and the intensity of the L-mode vanishes.
In this case, $n_\rL=0$ and the result in Table \ref{table:S=1-D} reduces to that in Table \ref{table:S=S} for $S=1$,
where $n_\rT$ plays the role of $n$ in Table \ref{table:S=1-D}.
There is a $x\leftrightarrow z$ correspondence between the two tables,
since the direction of the ordered moment is taken differently.

\begin{table}[t]
\caption{
Total moment sum rule obtained by extended spin-wave theory up to the $O(n)$ order for $S=1$ systems
with single-ion anisotropy of easy-plane type.
Components of the dynamical spin correlation function and the integrated intensities are shown.
Since the $x$-axis is taken along the ordered moment, the $xx$ component is for longitudinal spin fluctuation,
while the $yy$ and $zz$ components are for transverse one.
$S^{xx}(\bq,\omega)_{\rm 1-magnon}$ is for the L-mode in one-magnon process,
whereas $S^{xx}(\bq,\omega)_{\rm 2-magnon}$ is for the T-mode in two-magnon process.
The moment per one site is expressed as $|\langle S^x \rangle| = 2uv(1 - n_\rT - 2 n_\rL)$.
Here, $n_\rT= \langle a_{T}^\dagger a_{T} \rangle$ and $n_\rL=\langle a_{L}^\dagger a_{L} \rangle$.
The are calculated by the extended spin-wave theory as in Eq. (\ref{eqn:n-TL}).
$u$ and $v$ are defined by Eq. (\ref{eqn:uv-D}).
We can obtain $S(S+1)=2$ after adding all components of the intensity with the use of $u^2+v^2=1$.
}
\begin{tabular}{ccccc}
\hline
Component & Integrated intensity \\
\hline
$S^{xx}(\bq,\omega)_{\rm elastic}$ & $(2uv)^2(1-2n_\rT-4n_\rL)$ \\
$S^{xx}(\bq,\omega)_{\rm 1-magnon}$ & $(u^2-v^2)^2(1-n_\rT)$ \\
$S^{xx}(\bq,\omega)_{\rm 2-magnon}$ & $(2uv)^2(n_\rT+4n_\rL)$ \\
\hline
$S^{yy}(\bq,\omega)_{\rm 1-magnon}$ & $u^2(1-n_\rL)$ \\
$S^{yy}(\bq,\omega)_{\rm 2-magnon}$ & $v^2(n_\rT+n_\rL)$ \\
\hline
$S^{zz}(\bq,\omega)_{\rm 1-magnon}$ & $v^2(1-n_\rL)$ \\
$S^{zz}(\bq,\omega)_{\rm 2-magnon}$ & $u^2(n_\rT+n_\rL)$ \\
\hline
$[S^{yy}(\bq,\omega) + S^{zz}(\bq,\omega)]_{\rm 1-magnon}$ & $1 - n_\rL$ \\
$[S^{yy}(\bq,\omega) + S^{zz}(\bq,\omega)]_{\rm 2-magnon}$ & $n_\rT + n_\rL$ \\
\hline
$\sum_{\alpha=x,y,z}[S^{\alpha\alpha}(\bq,\omega)]_{\rm total}$ & $S(S+1)=2$ \\
\hline
\end{tabular}
\label{table:S=1-D}
\end{table}

\subsection{Calculated results}

\begin{figure}
\begin{center}
\includegraphics[width=6.5cm]{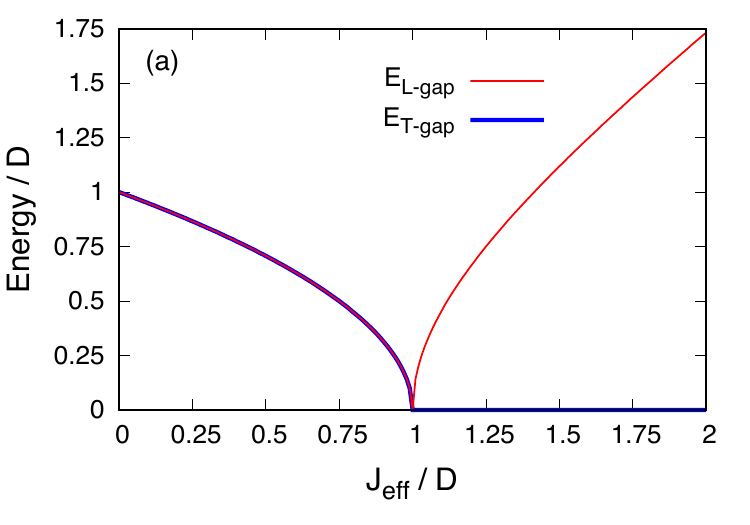}
\includegraphics[width=6.5cm]{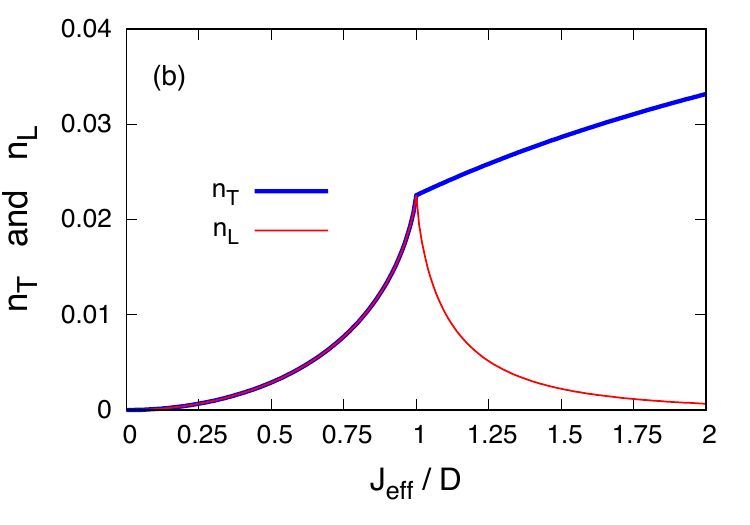}
\end{center}
\caption{
(Color online)
(a) $J_{\rm eff}/D$ dependence of the excitation gap in three-dimensional system.
$E_{\rm T-gap}$ and $E_{\rm L-gap}$ are for T- and L-modes, respectively.
They are given by $E_{\bk m}$ in Eq. (\ref{eqn:uv-Bogo-D}) with $\bk=(\pi,\pi,\pi)$.
In the disordered phase ($J_{\rm eff}/D<1$), notice that there is no distinction between the T- and L-modes.
(b) $J_{\rm eff}/D$ dependence of $n_{\rm T}=\braket{a_{i T}^\dagger a_{i T}}$ and $n_{\rm L}=\braket{a_{i L}^\dagger a_{i L}}$.
They are calculated with Eq. (\ref{eqn:n-TL}).
In the $J_{\rm eff}/D\rightarrow \infty$ limit, we obtain $(n_{\rm T},n_{\rm L})\rightarrow (0.0784,0)$.
}
\label{fig:gap-n}
\end{figure}

In this subsection, we use the result in Table \ref{table:S=1-D}
and resolve the integrated intensity into elastic, one-magnon, and two-magnon components.
We first show $J_{\rm eff}/D$ dependence of the excitation gap in Fig. \ref{fig:gap-n}(a).
The quantum critical point is located at $J_{\rm eff}/D=1$.
There are twofold degenerate excitation modes for $J_{\rm eff}/D<1$.
The excitation gap decreases with $J_{\rm eff}/D$ and becomes soft at $J_{\rm eff}/D=1$.
For $J_{\rm eff}/D>1$, the degeneracy is lifted by the emergence of the AF moment, along which we take the $x$-axis.
The excitation modes split into the T- and L-modes.
The T-mode stays gapless (Nambu-Goldstone mode),
whereas the L-mode acquires an excitation gap (Higgs amplitude mode) in the ordered phase.
We also show $n_{\rm T}$ and $n_{\rm L}$ in Fig. \ref{fig:gap-n}(b).
In the disordered phase, they increase with $J_{\rm eff}/D$ owing to the reduction of the excitation gap.
In the ordered phase, $n_{\rm T}$ increases and saturates in the $J_{\rm eff}/D\rightarrow \infty$ limit,
whereas $n_{\rm L}$ decreases with $J_{\rm eff}/D$ by the development of the excitation gap of the L-mode
and $n_{\rm L}$ vanishes in the $J_{\rm eff}/D\rightarrow \infty$ limit.

\begin{figure}
\begin{center}
\includegraphics[width=6.5cm]{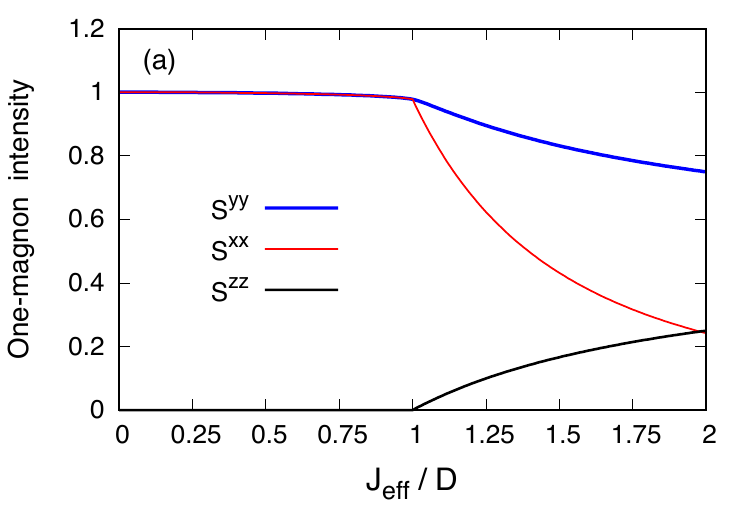}
\includegraphics[width=6.5cm]{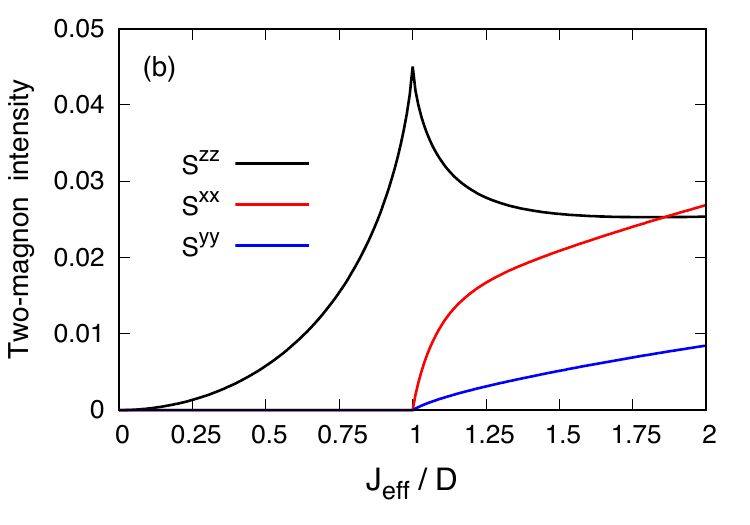}
\includegraphics[width=6.5cm]{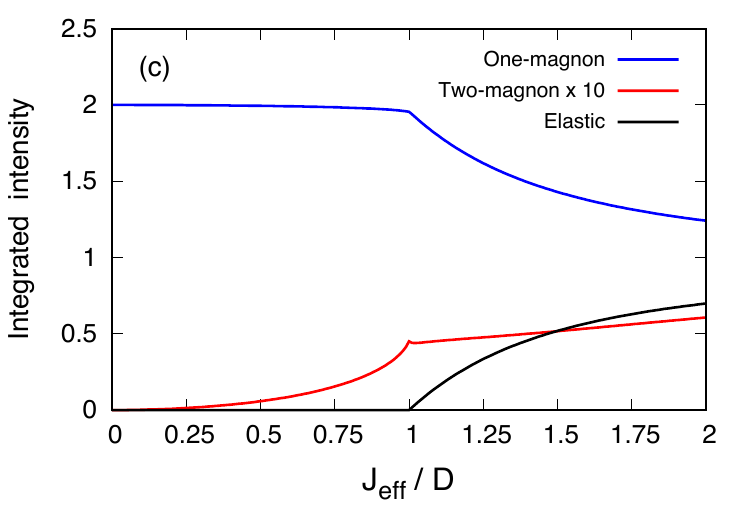}
\end{center}
\caption{
(Color online)
(a) $J_{\rm eff}/D$ dependence of the integrated intensity of the one-magnon component in three-dimensional system.
(b) Integrated intensity of the two-magnon component.
(c) Elastic, total one-magnon $[\sum_{\alpha=x,y,z} S^{\alpha\alpha}(\bq,\omega)_{\rm 1-magnon}]$,
and total two-magnon $[\sum_{\alpha=x,y,z} S^{\alpha\alpha}(\bq,\omega)_{\rm 2-magnon}]$ components
of the integrated intensity.
The elastic component only arises from $S^{xx}(\bq,\omega)$ (see Table \ref{table:S=1-D}).
Since the two-magnon component is small, the intensity is plotted by multiplying a factor of 10.
In the $J_{\rm eff}/D\rightarrow \infty$ limit, the three components of the integrated intensity become
(Elastic,~One-magnon,~Two-magnon)$\rightarrow(0.843,1,0.157)$.
}
\label{fig:S-3d}
\end{figure}

\begin{figure}
\begin{center}
\includegraphics[width=6.5cm]{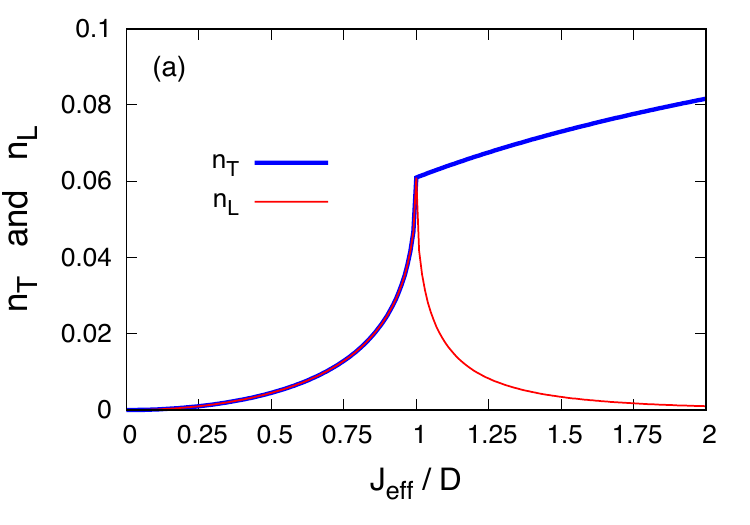}
\includegraphics[width=6.5cm]{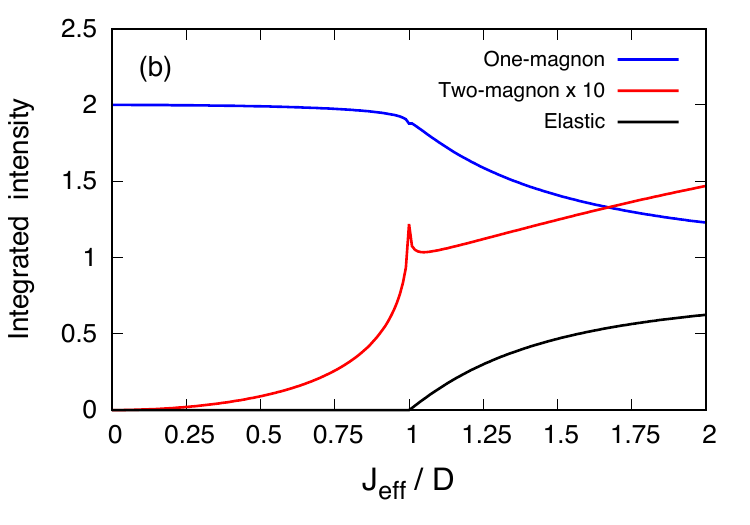}
\end{center}
\caption{
(Color online)
(a) $J_{\rm eff}/D$ dependence of $n_{\rm T}=\braket{a_{i T}^\dagger a_{i T}}$ and $n_{\rm L}=\braket{a_{i L}^\dagger a_{i L}}$
in two-dimensional system.
For the two-dimensional dispersion, we dropped the $\cos{k_z}$ term in $\gamma_\bk$ [see Eq. (\ref{eqn:ed-D})] and used $J_{\rm eff}=8J$.
In the $J_{\rm eff}/D\rightarrow \infty$ limit, we obtain $(n_{\rm T},n_{\rm L})\rightarrow (0.197,0)$.
(b) Elastic, total one-magnon, and total two-magnon components.
The two-magnon intensity is plotted by multiplying a factor of 10.
In the $J_{\rm eff}/D\rightarrow \infty$ limit, the three components of the integrated intensity become
(Elastic~,One-magnon~,Two-magnon)$\rightarrow(0.607,1,0.393)$.
}
\label{fig:S-2d}
\end{figure}

When we obtain $n_{\rm T}$ and $n_{\rm L}$, we can resolve the integrated intensity into elastic,
one-magnon, and two-magnon components, according to the results shown in Table \ref{table:S=1-D}.
Figure \ref{fig:S-3d}(a) shows the one-magnon component.
We can see that the intensity of $S^{zz}$ is suppressed by the strong easy-plane anisotropy,
and $S^{xx}$ and $S^{yy}$ carry the most intensity.
The intensities of $S^{xx}$ and $S^{yy}$ slightly decrease with $J_{\rm eff}/D$ in the disordered phase.
In the ordered phase, the longitudinal component $S^{xx}$ rapidly decreases,
whereas the transverse component $S^{yy}$ does not decrease so drastically.
In $S^{zz}$, a finite intensity appears in the ordered phase.
Figure \ref{fig:S-3d}(b) shows the two-magnon component.
In the disordered phase, the intensity of $S^{zz}$ develops with $J_{\rm eff}/D$ and takes a maximum value at $J_{\rm eff}/D=1$.
In the ordered phase, it decreases with $J_{\rm eff}/D$,
instead, finite intensities appear in $S^{xx}$ and $S^{yy}$ and they develop.
Figure \ref{fig:S-3d}(c) shows elastic, total one-magnon, and total two-magnon components of the integrated intensity.
The one-magnon intensity decreases with $J_{\rm eff}/D$, whereas the elastic intensity develops in the ordered phase.
The two-magnon intensity increases with $J_{\rm eff}/D$, showing a small peak at $J_{\rm eff}/D=1$.
We can see that the two-magnon intensity is very small compared with those of the elastic and one-magnon components
in the three-dimensional (3D) system.

For a two-dimensional (2D) system, on the other hand, the two-magnon intensity is enhanced as shown in Fig. \ref{fig:S-2d}.
Here, the $\cos(k_z)$ term in $\gamma_\bk$ [see Eq. (\ref{eqn:ed-D})] was dropped in the calculation.
In both 3D and 2D cases, we can see that the two-magnon intensity is enhanced in the vicinity of the quantum critical point.

In the $J_{\rm eff}/D\rightarrow \infty$ limit, the easy-plane anisotropy becomes negligible
and the result by the extended spin-wave theory reduces to that
by the conventional spin-wave theory for the isotropic Heisenberg model.
The extended spin-wave theory covers both the disordered and ordered phases.
It does not break down in 3D and 2D cases even in the vicinity of the quantum critical point.
In fact, it is known that the theory well explains the observed inelastic neutron spectra,
for instance, in \Tl,
\cite{Ruegg-2008}
\Ca,
\cite{Jain-2017}
\DLCB,
\cite{Hong-2017}
and \Cs
\cite{Hayashida-2019,Matsumoto-2020}
in the vicinity of the quantum critical point.

\section{$S=1/2$ spin dimer systems}

We study total moment sum rule in $S=1/2$ spin dimer systems.
As a typical example, we consider the following Hamiltonian for bilayer spin systems:
\begin{align}
\H = J_0 \sum_i  \bS_{il} \cdot \bS_{ir}
+ J \sum_{\langle i,j \rangle} \left( \bS_{il}\cdot \bS_{jl} + \bS_{ir}\cdot \bS_{jr} \right).
\label{eqn:H-dimer}
\end{align}
Here, $\bS_{il}$ and $\bS_{ir}$ are spin operator on the left and right side of a dimer at the $i$th site on a square lattice.
The summation $\sum_{\langle i,j \rangle}$ is taken over the nearest neighbor pairs on the square lattice.
$J_0(>0)$ and $J(>0)$ are AF exchange interaction parameters for intra- and inter-dimer interactions, respectively.

First, we introduce singlet ($|s\rangle$) and triplet [$|t_\alpha\rangle$ ($\alpha=x,y,z$)] states of a dimer
\cite{Sachdev-1990}
\begin{equation}
\begin{aligned}
&|s\rangle = \frac{1}{\sqrt{2}} ( |\uparrow\downarrow\rangle - |\downarrow\uparrow\rangle ),~~
|t_x\rangle = \frac{1}{\sqrt{2}} ( - |\uparrow\uparrow\rangle + |\downarrow\downarrow\rangle ), \cr
&|t_y\rangle = \frac{i}{\sqrt{2}} ( |\uparrow\uparrow\rangle + |\downarrow\downarrow\rangle ),~~
|t_z\rangle = \frac{1}{\sqrt{2}} ( |\uparrow\downarrow\rangle + |\downarrow\uparrow\rangle ).
\end{aligned}
\end{equation}
The three triplet states are expressed in the $x$, $y$, and $z$ representations,
as in the $p_x$, $p_y$, and $p_z$ orbital cases.

\subsection{Mean-field theory}

We take the $z$-axis along the ordered moment.
The mean-field Hamiltonian is then given by
\begin{align}
\H_{\rm MF} = J_0 \bS_{l} \cdot \bS_{r}
- 4J \left( S_{l}^z \langle S_{l}^z \rangle - S_{r}^z |\langle S_{r}^z \rangle| \right).
\label{eqn:H-mf-dimer}
\end{align}
In the presence of the staggered molecular field in the $z$ direction with respect to the left and right side of a dimer,
the threefold degeneracy of the triplet is lifted
and the singlet $|s\rangle$ and $|t_z\rangle$ triplet states are hybridized.
The local ground and excited states are given by the following form:
\cite{Matsumoto-2010}
\begin{equation}
\begin{aligned}
&|{\rm G}\rangle_i = u |s\rangle_i + v_i |t_z\rangle_i,~~~
|{\rm T}_x\rangle_i = | t_x \rangle_i, \cr
&|{\rm T}_y\rangle_i = | t_y \rangle_i,~~~~~~~~~~~~~~~~
|{\rm L}\rangle_i = -v_i |s\rangle_i + u |t_z\rangle_i.
\end{aligned}
\label{eqn:gs-dimer}
\end{equation}
Here, $|{\rm G}\rangle_i$ represents the ground state,
while $|{\rm T_x}\rangle_i$, $|{\rm T_y}\rangle_i$, and $|{\rm L}\rangle_i$ are excited states.
In Eq. (\ref{eqn:gs-dimer}),
\begin{align}
v_i=e^{i\bQ\cdot\br_i}v,
\end{align}
and $u$ and $v$ are real coefficients satisfying $u^2+v^2=1$.
$\br_i$ represents the position of the $i$th site, and $\bQ=(\pi,\pi,0)$ is the AF wave vector.
Thus, $e^{i\bQ\cdot\br_i}=\pm 1$ on the A and B sublattices, respectively.
In the disordered phase, $u=1$, $v=0$, $|{\rm G}\rangle_i=|s\rangle_i$, and $|{\rm L}\rangle_i=|t_z\rangle_i$.
The expectation values of the spin operators on the left and right side of a dimer are
\begin{align}
~_i\langle {\rm G}| \bS_{il} |{\rm G}\rangle_i = -~_i\langle {\rm G}| \bS_{ir} |{\rm G}\rangle_i = uv_i \be_z,
\end{align}
with $\be_z$ as a unit vector along the $z$ direction.
The ordered moment is staggered on the left and right side of a dimer.
It is also staggered on the A and B sublattices.
The expectation value in Eq. (\ref{eqn:H-mf-dimer}) is expressed as
\begin{align}
\langle S_{l}^z \rangle = - \langle S_{r}^z \rangle = uv.
\end{align}

The mean-field energy per one dimer is given by
\begin{align}
E_{\rm MF} &= \langle {\rm G}| [ J_0 \bS_l \cdot \bS_r - 2J (uv) (S_l^z-S_r^z) ] |{\rm G}\rangle \cr
&= ( J_0 - J_{\rm eff} ) v^2 + J_{\rm eff} v^4 - \frac{3}{4}J_0,
\end{align}
where 
\begin{align}
J_{\rm eff}=4J.
\end{align}
The coefficients $u$ and $v$ are determined so as to minimize the mean-field energy.
For $J_{\rm eff}/J_0 \ge 1$, they are determined as
\begin{align}
u = \sqrt{ \frac{1}{2}\left( 1 + \frac{J_0}{J_{\rm eff}} \right)},~~~~~~
v = \sqrt{ \frac{1}{2}\left( 1 - \frac{J_0}{J_{\rm eff}} \right)}.
\label{eqn:uv-dimer}
\end{align}
For $J_{\rm eff}/J_0 \le 1$, $u=1$ and $v=0$.
Therefore, $J_{\rm eff}=J_0$ represents a quantum critical point at which the disorder and ordered phases are separated.

\subsection{Extended spin-wave theory}

First, we introduce uniform and staggered components of a dimer by
\begin{align}
\bS_{i\pm} = \bS_{il} \pm \bS_{ir},
\end{align}
respectively.
Matrix form of $\bS_{i\pm}$ on the basis of the $|{\rm G}\rangle_i$, $|{\rm T_x}\rangle_i$, $|{\rm T_y}\rangle_i$, and $|{\rm L}\rangle_i$ states
are express as
\begin{align}
S_{i+}^x &=
\begin{pmatrix}
0 & 0 & iv_i & 0 \cr
0 & 0 & 0 & 0 \cr
-iv_i & 0 & 0 & -iu \cr
0 & 0 & iu & 0
\end{pmatrix} \cr
&\rightarrow i v_i ( a_{iy} - a_{iy}^\dagger ) - i u ( a_{iy}^\dagger a_{iL} - a_{iL}^\dagger a_{iy} ), \cr
S_{i-}^x &=
\begin{pmatrix}
0 & u & 0 & 0 \cr
u & 0 & 0 & -v_i \cr
0 & 0 & 0 & 0 \cr
0 & -v_i & 0 & 0
\end{pmatrix} \cr
&\rightarrow u ( a_{ix} + a_{ix}^\dagger ) - v_i ( a_{ix}^\dagger a_{iL} - a_{iL}^\dagger a_{ix} ), \cr
S_{i+}^y &=
\begin{pmatrix}
0 & -iv_i & 0 & 0 \cr
iv_i & 0 & 0 & iu \cr
0 & 0 & 0 & 0 \cr
0 & -iu & 0 & 0
\end{pmatrix} \cr
&\rightarrow - i v_i ( a_{ix} - a_{ix}^\dagger ) + i u ( a_{ix}^\dagger a_{iL} - a_{iL}^\dagger a_{ix} ), \cr
S_{i-}^y &=
\begin{pmatrix}
0 & 0 & u & 0\cr
0 & 0 & 0 & 0\cr
u & 0 & 0 & -v_i \cr
0 & 0 & -v_i & 0
\end{pmatrix}
\rightarrow u ( a_{iy} + a_{iy}^\dagger ) - v_i ( a_{iy}^\dagger a_{iL} - a_{iL}^\dagger a_{iy} ), \cr
S_{i+}^z &=
\begin{pmatrix}
0 & 0 & 0 & 0 \cr
0 & 0 & -i & 0\cr
0 & i & 0 & 0 \cr
0 & 0 & 0 & 0
\end{pmatrix}
\rightarrow -i ( a_{ix}^\dagger a_{iy} - a_{iy}^\dagger a_{ix} ), \cr
S_{i-}^z &=
\begin{pmatrix}
2uv_i & 0 & 0 & u^2-v^2\cr
0 & 0 & 0 & 0\cr
0 & 0 & 0 & 0 \cr
u^2-v^2 & 0 & 0 & -2uv_i
\end{pmatrix} \cr
&= 2uv_i {\bm{1}}
+ 
\begin{pmatrix}
0 & 0 & 0 & u^2-v^2\cr
0 & -2uv_i & 0 & 0\cr
0 & 0 & -2uv_i & 0 \cr
u^2-v^2 & 0 & 0 & -4uv_i
\end{pmatrix} \cr
&\rightarrow 2uv_i + (u^2-v^2) ( a_{iL} + a_{iL}^\dagger ) - 2uv_i a_{ix}^\dagger a_{ix} \cr
&~~~
- 2uv_i a_{iy}^\dagger a_{iy} - 4uv_i a_{iL}^\dagger a_{iL}.
\label{eqn:S-dimer}
\end{align}
Here, we introduced bosons ($a_{iG}$, $a_{ix}$, $a_{iy}$, $a_{iL}$)
for the ($|{\rm G}\rangle_i$, $|{\rm T}_x\rangle_i$, $|{\rm T}_y\rangle_i$, $|{\rm G}\rangle_i$) states, respectively.
The bosons are subjected to the local constraint
\begin{align}
a_{iG}^\dagger a_{iG} + a_{ix}^\dagger a_{ix}+ a_{iy}^\dagger a_{iy} + a_{iL}^\dagger a_{iL} = 1.
\label{eqn:constraint-dimer}
\end{align}
With the use of the local constraint, we can express
\begin{equation}
\begin{aligned}
&a_{iG} \rightarrow \left( 1 - a_{ix}^\dagger a_{ix}+ a_{iy}^\dagger a_{iy} + a_{iL}^\dagger a_{iL} \right)^{\frac{1}{2}}, \cr
&a_{iG}^\dagger \rightarrow \left( 1 - a_{ix}^\dagger a_{ix}+ a_{iy}^\dagger a_{iy} + a_{iL}^\dagger a_{iL} \right)^{\frac{1}{2}}.
\label{eqn:replace-dimer}
\end{aligned}
\end{equation}
Equations (\ref{eqn:constraint-dimer}) and (\ref{eqn:replace-dimer}) were used in Eq. (\ref{eqn:S-dimer}),
and we retained up to quadratic order of bosons.
The spin operators $\bS_{il}$ and $\bS_{ir}$ are then given by
\begin{align}
\bS_{il} = \frac{1}{2} \left( \bS_{i+} + \bS_{i-} \right),~~~~~~
\bS_{ir} = \frac{1}{2} \left( \bS_{i+} - \bS_{i-} \right).
\end{align}

For the intra-dimer interaction, $\bS_{il}\cdot\bS_{ir}$ is expressed as
\begin{align}
\bS_{il}\cdot\bS_{ir} &=
\begin{pmatrix}
\frac{1}{4}(-3u^2+v^2) & 0 & 0 & uv_i \cr
0 & \frac{1}{4} & 0 & 0\cr
0 & 0 & \frac{1}{4} & 0 \cr
uv_i & 0 & 0 & \frac{1}{4}(u^2-3v^2)
\end{pmatrix} \cr
&= \frac{1}{4}(-3u^2+v^2) {\bm{1}} \cr
&~~~
+
{\footnotesize
\begin{pmatrix}
0 & 0 & 0 & uv_i \cr
0 & \frac{1}{4}(1+3u^2-v^2) & 0 & 0\cr
0 & 0 & \frac{1}{4}(1+3u^2-v^2) & 0 \cr
uv_i & 0 & 0 & u^2-v^2
\end{pmatrix}
}
\cr
&= -\frac{3}{4} + v^2 + u^2 \left( a_{ix}^\dagger a_{ix} + a_{iy}^\dagger a_{iy} \right) \cr
&~~~
+ (u^2-v^2) a_{iL}^\dagger a_{iL} + uv_i \left( a_{iL} + a_{iL}^\dagger \right).
\label{eqn:intra}
\end{align}

\begin{figure}
\begin{center}
\includegraphics[width=5cm]{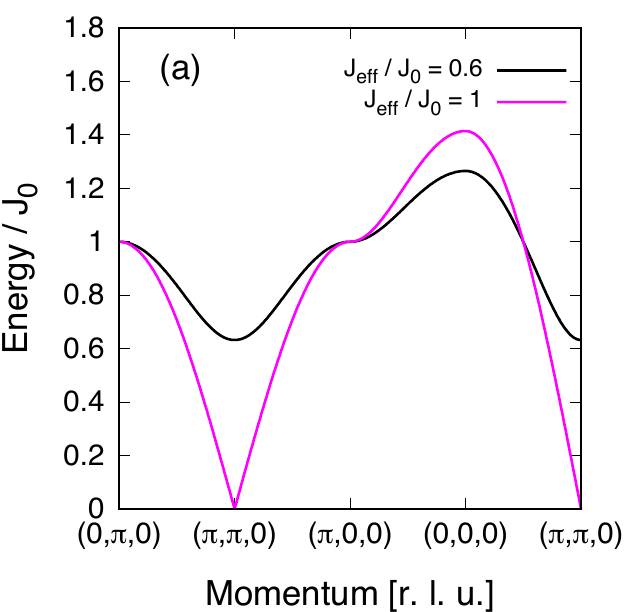}
\includegraphics[width=5cm]{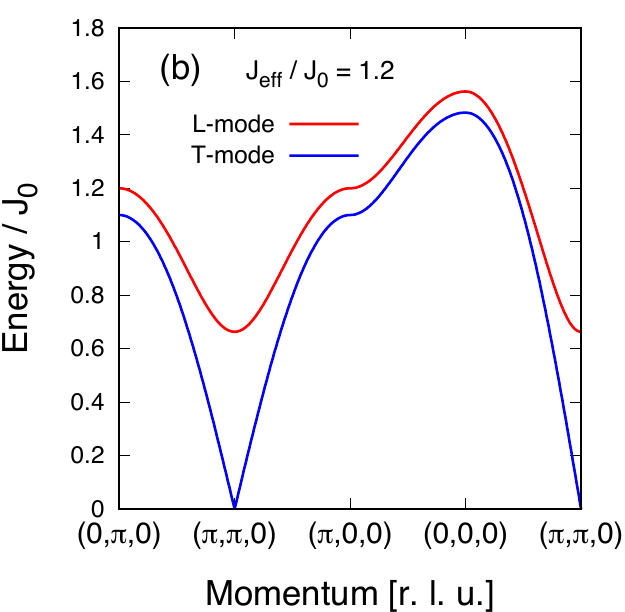}
\end{center}
\caption{
(Color online)
Magnon dispersion relation in $S=1/2$ spin dimer systems on a bilayer square lattice.
(a) For $J_{\rm eff}/J_0\le 1$ (disordered phase).
(b) For $J_{\rm eff}/J_0 > 1$ (ordered phase).
The excitation modes split into single L- and twofold T-modes.
}
\label{fig:dimer-mag}
\end{figure}

Substituting Eqs. (\ref{eqn:S-dimer}) and (\ref{eqn:intra}) into Eq. (\ref{eqn:H-dimer}),
we obtain the following form of the Hamiltonian:
\cite{Matsumoto-2010}
\begin{align}
\H = \sum_\bk &\sum_{m=x,y,L} \left[ \epsilon_{\bk m} a_{\bk m}^\dagger a_{\bk m}
+ \frac{1}{2} \Delta_{\bk m} ( a_{\bk m} a_{\bk m} + a_{\bk m}^\dagger a_{\bk m}^\dagger ) \right],
\label{eqn:H-dimer-k}
\end{align}
where
\begin{equation}
\begin{aligned}
&\epsilon_{\bk x} = \epsilon_{\bk y} = u^2 J_0 + 2(uv)^2 J_{\rm eff} + (u^2-v^2)\gamma_\bk, \cr
&\epsilon_{\bk L} = (u^2-v^2)J_0 + 4(uv)^2 J_{\rm eff} + (u^2-v^2)^2 \gamma_\bk, \cr
&\Delta_{\bk x} = \Delta_{\bk y} = \gamma_\bk, \cr
&\Delta_{\bk L} = (u^2-v^2)^2 \gamma_\bk, \cr
&\gamma_\bk = J ( \cos{k_x} + \cos{k_y} ).
\end{aligned}
\label{eqn:ed-dimer}
\end{equation}
Here, $a_{\bk m}$ is the Fourier transformed operator of $a_{im}$.
Notice that Eq. (\ref{eqn:ed-dimer}) is essentially the same as Eq. (\ref{eqn:ed-D}).
The energy of the excitation mode and the Bogoliubov transformation are the same as in Eq. (\ref{eqn:uv-Bogo-D}).
The expectation value is calculated as
\begin{align}
n_m = \langle a_{im}^\dagger a_{im} \rangle
= \frac{1}{N} \sum_\bk v_{\bk m}^2~~~(m=x,y,L).
\label{eqn:n-xyL}
\end{align}
For $J_{\rm eff}/J_0<1$, the disordered phase is realized.
The magnon mode is threefold degenerate and has a finite excitation gap at the AF wave vector $\bQ=(\pi,\pi,0)$,
as shown in Fig. \ref{fig:dimer-mag}(a).
At the quantum critical point, $J_{\rm eff}/J_0 = 1$, the excitation becomes soft
showing a linear dispersion relation around $\bk=\bQ$.
For $J_{\rm eff}/J_0>1$, the ordered phase is stabilized
and the threefold degenerate modes split into single L- and twofold  T-modes.
The T-modes are gapless, whereas the L-mode is gapped, as shown in Fig. \ref{fig:dimer-mag}(b).

\subsection{Total moment sum rule}

For spin dimers, the integrated dynamical spin correlation function is expressed as
\begin{align}
&\frac{1}{N} \sum_i \sum_{\alpha=x,y,z} \frac{1}{2} \langle S_{il}^\alpha S_{il}^\alpha + S_{ir}^\alpha S_{ir}^\alpha \rangle \cr
&~~~
=
\frac{1}{N} \sum_i \sum_{\alpha=x,y,z} \frac{1}{4} \langle S_{i+}^\alpha S_{i+}^\alpha + S_{i-}^\alpha S_{i-}^\alpha \rangle.
\label{eqn:integrated-S-dimer}
\end{align}
Here, $N$ is number of dimer sites.

\subsubsection{Ordered moment}

From Eq. (\ref{eqn:S-dimer}), we can calculate the expectation value of the ordered moment per site.
Since the moment is staggered on the left and right side of a dimer, it is given by
\begin{align}
\frac{1}{2} \left| \langle S_{il}^z \rangle - \langle S_{ir}^z \rangle \right|
= \frac{1}{2} | \langle S_{i-}^z \rangle |
= uv( 1 - n_x - n_y - 2 n_\rL ).
\label{eqn:Sz-av-dimer}
\end{align}
where we introduced
\begin{align}
n_x = \langle a_{ix}^\dagger a_{ix} \rangle,~~~
n_y = \langle a_{iy}^\dagger a_{iy} \rangle,~~~
n_\rL = \langle a_{iL}^\dagger a_{iL} \rangle.
\end{align}
In Eq. (\ref{eqn:Sz-av-dimer}), $uv$ is the mean-field value,
whereas the factor $(1 - n_x - n_y - 2n_\rL)$ is from the quantum correction.
Notice that the moment already shrinks in the mean-field level,
owing to the dimerization by the AF intra-dimer interaction.

\subsubsection{$z$ component}

From Eq. (\ref{eqn:S-dimer}), $(S_{i+}^z)^2$ is expressed as
\begin{align}
(S_{i+}^z)^2 =
\begin{pmatrix}
0 & 0 & 0 & 0\cr
0 & 1 & 0 & 0\cr
0 & 0 & 1 & 0 \cr
0 & 0 & 0 & 0
\end{pmatrix}
= a_{ix}^\dagger a_{ix} + a_{iy}^\dagger a_{iy}.
\end{align}
Since the first and second terms are from
$a_{ix}^\dagger a_{iy} a_{iy}^\dagger a_{ix}$ and $a_{iy}^\dagger a_{ix} a_{ix}^\dagger a_{iy}$ processes, respectively,
they are two-magnon process.
For Eq. (\ref{eqn:integrated-S-dimer}), we then obtain
\begin{align}
\langle S_{i+}^z S_{i+}^z \rangle_{\rm 2-magnon}
= \langle a_{ix}^\dagger a_{ix} \rangle + \langle a_{iy}^\dagger a_{iy} \rangle
= n_x + n_y.
\end{align}
In the same way, $(S_{i-}^z)^2$ is expressed as
\begin{align}
(S_{i-}^z)^2 &= (2uv)^2 {\bm{1}} \cr
&~~~
+ 2(2uv_i)
\begin{pmatrix}
0 & 0 & 0 & u^2-v^2\cr
0 & -2uv_i & 0 & 0\cr
0 & 0 & -2uv_i & 0 \cr
u^2-v^2 & 0 & 0 & -4uv_i
\end{pmatrix} \cr
&~~~
+
\begin{pmatrix}
0 & 0 & 0 & u^2-v^2\cr
0 & -2uv_i & 0 & 0\cr
0 & 0 & -2uv_i & 0 \cr
u^2-v^2 & 0 & 0 & -4uv_i
\end{pmatrix}^2.
\end{align}
Here, the first two terms are elastic component, while the last term is inelastic one.
The last term can be expressed as
\begin{align}
&
\begin{pmatrix}
0 & 0 & 0 & u^2-v^2\cr
0 & -2uv_i & 0 & 0\cr
0 & 0 & -2uv_i & 0 \cr
u^2-v^2 & 0 & 0 & -4uv_i
\end{pmatrix}^2 \cr
&=
{\footnotesize
\begin{pmatrix}
(u^2-v^2)^2 & 0 & 0 & -4uv_i(u^2-v^2) \cr
0 & (2uv)^2 & 0 & 0 \cr
0 & 0 & (2uv)^2 & 0 \cr
-4uv_i(u^2-v^2) & 0 & 0 & (u^2-v^2)^2 + (4uv)^2
\end{pmatrix}
}
\cr
&\xrightarrow{{\rm DP}}
(u^2-v^2)^2 a_{iG}^\dagger a_{iG} + (2uv)^2a_{ix}^\dagger a_{ix} \cr
&~~~
+ (2uv)^2 a_{iy}^\dagger a_{iy} + (u^2-v^2)^2 a_{iL}^\dagger a_{iL} + (4uv)^2 a_{iL}^\dagger a_{iL}.
\label{eqn:Sz-dimer2}
\end{align}
In Eq. (\ref{eqn:Sz-dimer2}), the first term is from $a_{iG}^\dagger a_{iL} a_{iL}^\dagger a_{iG}$ and is one-magnon process.
The expectation value is calculated as
$\langle a_{iG}^\dagger a_{iG}\rangle = \langle 1 - a_{ix}^\dagger a_{ix} - a_{iy}^\dagger a_{iy} - a_{iL}^\dagger a_{iL} \rangle = 1 - n_x - n_y - n_\rL$.
The second term is from $a_{ix}^\dagger a_{ix} a_{ix}^\dagger a_{ix}$ and is two-magnon process.
The third term is from $a_{iy}^\dagger a_{iy} a_{iy}^\dagger a_{iy}$ and is two-magnon process.
The fourth term is from $a_{iL}^\dagger a_{iG} a_{iG}^\dagger a_{iL}$ and is one-magnon process.
The last term is from $a_{iL}^\dagger a_{iL} a_{il}^\dagger a_{iL}$ and is two-magnon process.
Thus, the expectation value is resolved as
\begin{align}
&\langle S_{i-}^z S_{i-}^z \rangle_{\rm elastic} = (2uv)^2 - 2(2uv)^2 ( n_x + n_y + 2n_\rL ), \cr
&\langle S_{i-}^z S_{i-}^z \rangle_{\rm 1-magnon} = (u^2-v^2)^2 ( 1 - n_x - n_y ) , \\
&\langle S_{i-}^z S_{i-}^z \rangle_{\rm 2-magnon} = (2uv)^2 ( n_x + n_y + 4 n_\rL ).
\nonumber
\end{align}

\subsubsection{$x$ component}

From Eq. (\ref{eqn:S-dimer}), $(S_{i+}^x)^2$ is expressed as
\begin{align}
(S_{i+}^x)^2 &=
\begin{pmatrix}
v^2 & 0 & 0 & uv_i \cr
0 & 0 & 0 & 0\cr
0 & 0 & u^2+v^2 & 0 \cr
uv_i & 0 & 0 & u^2
\end{pmatrix} \cr
&\xrightarrow{\rm DP}
v^2 a_{iG}^\dagger a_{iG} +  v^2 a_{iy}^\dagger a_{iy} + u^2 a_{iy}^\dagger a_{iy} + u^2 a_{iL}^\dagger a_{iL}.
\label{eqn:Spx}
\end{align}
The first term is from $a_{iG}^\dagger a_{iy} a_{iy}^\dagger a_{iG}$ and is one-magnon process.
The second term is from $a_{iy}^\dagger a_{iG} a_{iG}^\dagger a_{iy}$ and is one-magnon process.
The third term is from $a_{iy}^\dagger a_{iL} a_{iL}^\dagger a_{iy}$ and is two-magnon process.
The last term is from $a_{iL}^\dagger a_{iy} a_{iy}^\dagger a_{iL}$ and is two-magnon process.
Thus, the expectation value is resolved as
\begin{equation}
\begin{aligned}
&\langle S_{i+}^x S_{i+}^x \rangle_{\rm 1-magnon} = v^2 ( 1 - n_x - n_\rL ), \cr
&\langle S_{i+}^x S_{i+}^x \rangle_{\rm 2-magnon} = u^2 ( n_y + n_\rL ).
\end{aligned}
\end{equation}

From Eq. (\ref{eqn:S-dimer}), $(S_{i-}^x)^2$ is expressed as
\begin{align}
(S_{i-}^x)^2 &=
\begin{pmatrix}
u^2 & 0 & 0 & -uv_i \cr
0 & u^2+v^2 & 0 & 0\cr
0 & 0 & 0 & 0 \cr
-uv_i & 0 & 0 & v^2
\end{pmatrix} \cr
&\xrightarrow{\rm DP}
u^2 a_{iG}^\dagger a_{iG} +  u^2 a_{ix}^\dagger a_{ix} + v^2 a_{ix}^\dagger a_{ix} + v^2 a_{iL}^\dagger a_{iL}.
\label{eqn:Snx}
\end{align}
The first term is from $a_{iG}^\dagger a_{ix} a_{ix}^\dagger a_{iG}$ and is one-magnon process.
The second term is from $a_{ix}^\dagger a_{iG} a_{iG}^\dagger a_{ix}$ and is one-magnon process.
The third term is from $a_{ix}^\dagger a_{iL} a_{iL}^\dagger a_{ix}$ and is two-magnon process.
The last term is from $a_{iL}^\dagger a_{ix} a_{ix}^\dagger a_{iL}$ and is two-magnon process.
Thus, the expectation value is resolved as
\begin{equation}
\begin{aligned}
&\langle S_{i-}^x S_{i-}^x \rangle_{\rm 1-magnon} = u^2 ( 1 - n_y - n_\rL ), \cr
&\langle S_{i-}^x S_{i-}^x \rangle_{\rm 2-magnon} = v^2 ( n_x + n_\rL ).
\end{aligned}
\end{equation}

\subsubsection{$y$ component}

From Eq. (\ref{eqn:S-dimer}), $(S_{i\pm}^y)^2$ are expressed as
\begin{equation}
\begin{aligned}
(S_{i+}^y)^2 &=
\begin{pmatrix}
v^2 & 0 & 0 & uv_i \cr
0 & u^2+v^2 & 0 & 0\cr
0 & 0 & 0 & 0 \cr
uv_i & 0 & 0 & u^2
\end{pmatrix}, \cr
(S_{i+}^y)^2 &=
\begin{pmatrix}
u^2 & 0 & 0 & -uv_i \cr
0 & 0 & 0 & 0\cr
0 & 0 & u^2+v^2 & 0 \cr
-uv_i & 0 & 0 & v^2
\end{pmatrix}.
\label{eqn:Spny}
\end{aligned}
\end{equation}
Comparing Eq. (\ref{eqn:Spny}) with Eqs. (\ref{eqn:Spx}) and (\ref{eqn:Snx}), we notice that the role of $x$ and $y$ are interchanged.
Therefore, we obtain
\begin{equation}
\begin{aligned}
&\langle S_{i+}^y S_{i+}^y \rangle_{\rm 1-magnon} = v^2 ( 1 - n_y - n_\rL ), \cr
&\langle S_{i+}^y S_{i+}^y \rangle_{\rm 2-magnon} = u^2 ( n_x + n_\rL ), \cr
&\langle S_{i-}^y S_{i-}^y \rangle_{\rm 1-magnon} = u^2 ( 1 - n_x - n_\rL ), \cr
&\langle S_{i-}^y S_{i-}^y \rangle_{\rm 2-magnon} = v^2 ( n_y + n_\rL ).
\end{aligned}
\end{equation}

Substituting the above results in Eq. (\ref{eqn:integrated-S-dimer}), we resolve the integrated correlation function.
The result for the total moment sum rule in summarized in Table \ref{table:S=1/2-dimer}.
In the ordered phase far from the quantum critical point, $u=v=1/\sqrt{2}$.
In this case, the one-magnon intensity of the L-mode vanishes
and the result in Table \ref{table:S=1/2-dimer} reduces to that in Table \ref{table:S=S} for S = 1/2,
where $n_\rL=0$ and $n_x$ and $n_y$ play the role of $n$ in Table \ref{table:S=S}.

\begin{table}[t]
\caption{
Total moment sum rule obtained by the extended spin-wave theory up to the $O(n)$ order for $S=1/2$ dimer systems.
Components of the dynamical spin correlation function and the integrated intensities are shown.
Since the $z$-axis is taken along the ordered moment, the $zz$ component is for longitudinal spin fluctuation,
while the $xx$ and $yy$ components are for transverse one.
$S^{zz}(\bq,\omega)_{\rm 1-magnon}$ is for the L-mode in one-magnon process,
whereas $S^{zz}(\bq,\omega)_{\rm 2-magnon}$ is for the T-mode in two-magnon process.
The moment per one site is expressed as $\langle S^z \rangle = uv(1 - n_x - n_y - 2n_\rL)$.
Here, $n_x= \langle a_{x}^\dagger a_{x} \rangle$, $n_y= \langle a_{y}^\dagger a_{y} \rangle$,
and $n_\rL=\langle a_{L}^\dagger a_{L} \rangle$.
They are calculated by the extended spin-wave theory as in Eq. (\ref{eqn:n-xyL}).
$u$ and $v$ are defined by Eq. (\ref{eqn:uv-dimer}).
We can obtain $S(S+1)=3/4$ after adding all components with the use of $u^2+v^2=1$.
}
\begin{tabular}{ccccc}
\hline
Component & Integrated intensity \\
\hline
$S^{zz}(\bq,\omega)_{\rm elastic}$ & $(uv)^2 - 2(uv)^2 (n_x + n_y + 2n_\rL)$ \\
$S^{zz}(\bq,\omega)_{\rm 1-magnon}$ & $\frac{1}{4}(u^2-v^2)^2(1 - n_x - n_y)$ \\
$S^{zz}(\bq,\omega)_{\rm 2-magnon}$ & $(uv)^2(n_x + n_y + 4n_\rL) + \frac{1}{4}( n_x + n_y )$ \\
\hline
$S^{xx}(\bq,\omega)_{\rm 1-magnon}$ & $\frac{1}{4} - \frac{1}{4}v^2 (n_x+n_\rL) - \frac{1}{4}u^2 (n_y + n_\rL )$ \\
$S^{xx}(\bq,\omega)_{\rm 2-magnon}$ & $\frac{1}{4} v^2 (n_x + n_\rL) + \frac{1}{4} u^2 (n_y + n_\rL )$ \\
\hline
$S^{yy}(\bq,\omega)_{\rm 1-magnon}$ & $\frac{1}{4} - \frac{1}{4} u^2 (n_x+n_\rL) - \frac{1}{4} v^2 (n_y + n_\rL )$ \\
$S^{yy}(\bq,\omega)_{\rm 2-magnon}$ & $\frac{1}{4} u^2 (n_x + n_\rL) + \frac{1}{4} v^2 (n_y + n_\rL )$ \\
\hline
$[S^{xx}(\bq,\omega)+S^{yy}(\bq,\omega)]_{\rm 1-magnon}$ & $\frac{1}{2} - \frac{1}{4} ( n_x + n_y + 2n_\rL )$ \\
$[S^{xx}(\bq,\omega)+S^{yy}(\bq,\omega)]_{\rm 2-magnon}$ & $\frac{1}{4} ( n_x + n_y + 2n_\rL )$ \\
\hline
$\sum_{\alpha=x,y,z}[S^{\alpha\alpha}(\bq,\omega)]_{\rm total}$ & $S(S+1)=\frac{3}{4}$ \\
\hline
\end{tabular}
\label{table:S=1/2-dimer}
\end{table}

\subsubsection{Application to \Cu}

\Cu~is a dimerized quasi one-dimensional system with $J_0=0.442$ meV and $J=0.106$ meV for intra- and inter-dimer interactions.
\cite{Tennant-2003}
This copper nitrate does not show magnetic ordering even at low temperatures.
In the disordered phase, $u=1$ and $v=0$ and there is a finite excitation gap.
Notice that the theory works in one-dimensional systems with a finite excitation gap.
\cite{Gopalan-1994}
The integrated intensity of one-magnon and two-magnon processes are expressed as
\begin{equation}
\begin{aligned}
&I_{\rm 1-magnon}^{\rm total} = \frac{3}{4} - \frac{1}{2}( n_x + n_y + n_z ), \cr
&I_{\rm 2-magnon}^{\rm total} = \frac{1}{2}( n_x + n_y + n_z ).
\end{aligned}
\end{equation}
Here, we rewrite $n_\rL\rightarrow n_z=\langle t_z^\dagger t_z \rangle$,
since $|t_L\rangle=|t_z\rangle$ in the disordered phase [see Eq. (\ref{eqn:gs-dimer})].
For the isotropic exchange interactions, the triplet excitation is threefold degenerate
and $\epsilon_{k m}$, $\Delta_{k m}$, and $E_{k m}$ ($m=x,y,z$) are given by
\cite{Gopalan-1994}
\begin{align}
&\epsilon_{k m} = J_0 - \frac{1}{2} J \cos{k}, \cr
&\Delta_{k m} = - \frac{1}{2} J \cos{k}, \\
&E_{k m} = \sqrt{\epsilon_k^2 - \Delta_k^2 } = \sqrt{ J_0 ( J_0 - J \cos{k} ) }.
\nonumber
\end{align}
The values of $n(\equiv n_x=n_y=n_z)$ is calculated as
\begin{align}
n = \frac{1}{2\pi} \int_{-\pi}^\pi dk \frac{1}{2} \left( \frac{\epsilon_{k m}}{E_{k m}} - 1 \right)
\simeq 0.00187.
\end{align}
The value of $n$ is strongly reduced by the excitation gap.
The ratio of the integrated intensity for the two-magnon and one-magnon processes is estimated as
\begin{align}
\frac{I_{\rm 2-magnon}^{\rm total}}{I_{\rm 1-magnon}^{\rm total}} \simeq 0.00376.
\end{align}
Therefore, the two-magnon intensity is quite weak compared to that for the one-magnon.
This is consistent with the observed result that the ratio is of the order of $10^{-2}$ in \Cu.
\cite{Tennant-2003}

\begin{table*}[t]
\caption{
Total moment sum rule for $S=3/2$ spin dimer systems.
Components of the dynamical spin correlation function and the integrated intensities are shown.
Since the $z$-axis is taken along the ordered moment, the $zz$ component is for longitudinal spin fluctuation,
while the $xx$ and $yy$ components are for transverse one.
$S^{zz}(\bq,\omega)_{\rm 1-magnon}$ is for the L-mode in one-magnon process,
whereas $S^{zz}(\bq,\omega)_{\rm 2-magnon}$ is for the T-mode in two-magnon process.
The moment per one site is expressed as $\langle S^z \rangle = \sqrt{5}uv(1 - n_x - n_y - 2n_\rL)$.
Here, $n_x= \langle a_{x}^\dagger a_{x} \rangle$, $n_y= \langle a_{y}^\dagger a_{y} \rangle$,
and $n_\rL=\langle a_{L}^\dagger a_{L} \rangle$.
Notice that the value of $S(S+1)=15/4$ is not obtained after adding all components,
since we restricted the low-energy singlet and triplet states and discarded the other high-energy states.
}
\begin{tabular}{ccccc}
\hline
Component & Integrated intensity \\
\hline
$[S^{zz}(\bq,\omega)]_{\rm elastic}$ & $5(uv)^2-2\sqrt{5}(uv)^2(n_x+n_y+2n_\rL)$ \\
$[S^{zz}(\bq,\omega)]_{\rm 1-magnon}$ & $\frac{5}{4}(u^2-v^2)^2 (1-n_x-n_y)$ \\
$[S^{zz}(\bq,\omega)]_{\rm 2-magnon}$ & $5(uv)^2(n_x+n_y+4n_\rL) + \frac{1}{4}(n_x+n_y)$ \\
\hline
$[S^{xx}(\bq,\omega)]_{\rm 1-magnon}$ & $\left(\frac{5}{4}u^2+\frac{1}{4}v^2\right)
                                                                                - \frac{1}{4}v^2 n_x - \frac{5}{4}u^2 n_y - \left( \frac{5}{4}u^2+\frac{1}{4}v^2\right)n_\rL$ \\
$[S^{xx}(\bq,\omega)]_{\rm 2-magnon}$ & $\frac{5}{4}v^2 n_x + \frac{1}{4}u^2 n_y + \left(\frac{1}{4}u^2+\frac{5}{4}v^2\right) n_\rL$ \\
$[S^{yy}(\bq,\omega)]_{\rm 1-magnon}$ & $\left(\frac{5}{4}u^2+\frac{1}{4}v^2\right)
                                                                                - \frac{5}{4}u^2 n_x - \frac{1}{4}v^2 n_y - \left( \frac{5}{4}u^2+\frac{1}{4}v^2\right)n_\rL$ \\
$[S^{yy}(\bq,\omega)]_{\rm 2-magnon}$ & $\frac{1}{4}u^2 n_x + \frac{5}{4}v^2 n_y + \left(\frac{1}{4}u^2+\frac{5}{4}v^2\right) n_\rL$ \\
\hline
$[S^{xx}(\bq,\omega)+S^{yy}(\bq,\omega)]_{\rm 1-magnon}$ & $ \left( \frac{5}{4}u^2+\frac{1}{4}v^2\right)(2-n_x-n_y-2n_\rL)$ \\
$[S^{xx}(\bq,\omega)+S^{yy}(\bq,\omega)]_{\rm 2-magnon}$ & $\left(\frac{1}{4}u^2+\frac{5}{4}v^2\right) (n_x + n_y + 2 n_\rL )$ \\
\hline
\end{tabular}
\label{table:S=3/2-dimer}
\end{table*}

\subsection{$S=3/2$ spin dimer case}

For a $S=3/2$ dimer, there are $16(=4\times 4)$ local states.
The low-energy levels are formed by singlet and triplet states, while there are other high-energy 12 states.
The high-energy modes have a quite weak intensity for inelastic neutron scattering,
and we restrict the low-energy singlet and triplet states and discard the other high-energy states.
In this case, the spin operators are expressed in a $4\times 4$ matrix form.
The characteristic point of the matrix elements of the spin operators is that
the element between the singlet and triplet states is enhanced by a factor of $\sqrt{5}$.
This enhancement factor appears in the staggered component of the spin operator for a dimer,
i.e. $\bS_{i-}\rightarrow \sqrt{5}\bS_{i-}$ form the value of for the $S=1/2$ dimer.
The other matrix elements between triplet states are unchanged.
This means that the uniform component does not change, i.e. $\bS_{i+}\rightarrow \bS_{i+}$.
Therefore, we can discuss the total moment sum rule in the $S=3/2$ dimer case in parallel with that for the $S=1/2$ dimer.
We summarize the result for the $S=3/2$ dimer in Table \ref{table:S=3/2-dimer}.

\subsubsection{Application to $S=3/2$ dimer system \Cr}

\begin{figure}
\begin{center}
\includegraphics[width=5cm]{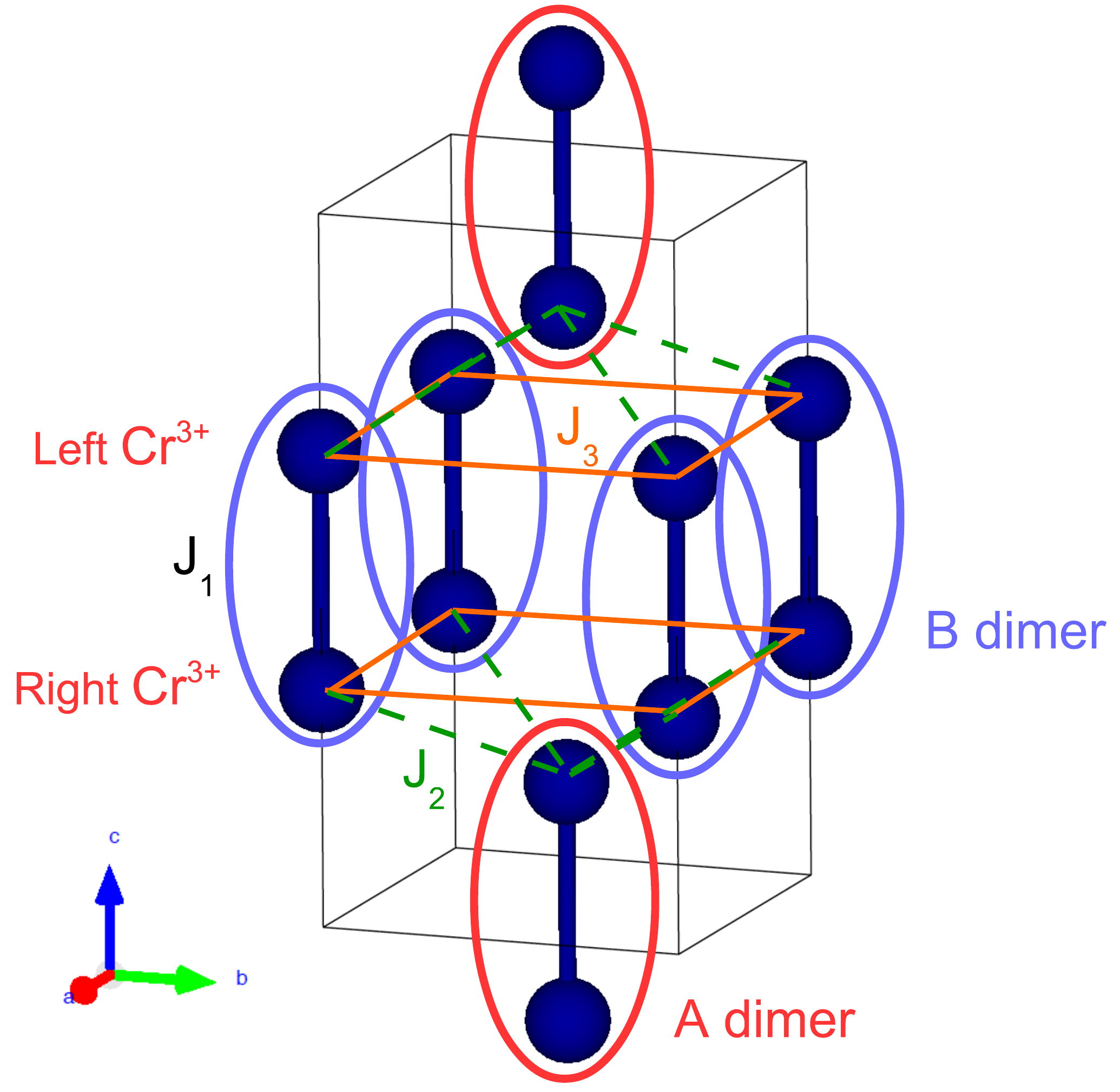}
\end{center}
\caption{
(Color online)
Crystal structure of \Cr.
$J_1$ is an intra-dimer interaction, while $J_2$ and $J_3$ are inter-dimer interactions.
$J_2$ connects dimers along the $c$ direction.
$J_3$ connects dimers in the $ab$ plane.
The values are estimated as
$J_1= 5.25$ meV, $J_2= -0.475$ meV, $J_3= -0.1$ meV.
The AF wave vector is $\bQ=(0,0,2\pi)$ in the reciprocal lattice unit.
}
\label{fig:dimer}
\end{figure}

\begin{figure}
\begin{center}
\includegraphics[width=5cm]{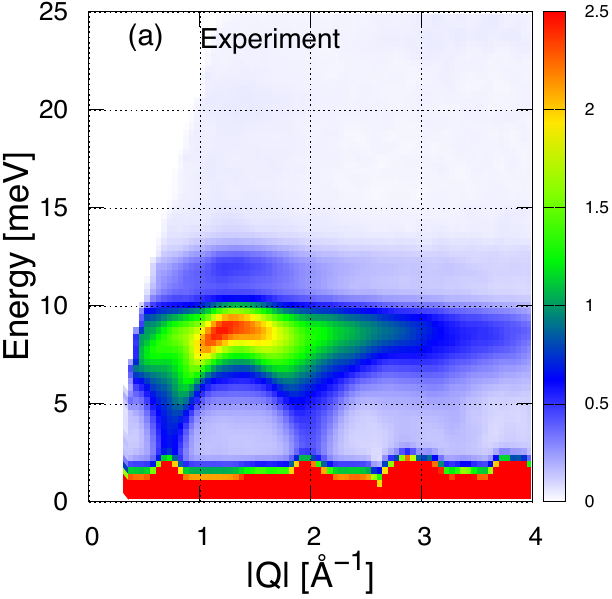}
\includegraphics[width=5cm]{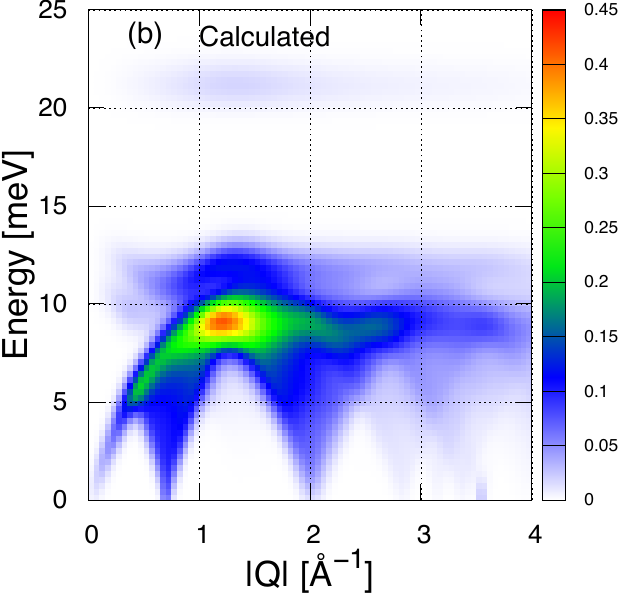}
\end{center}
\caption{
(Color online)
The intensity of inelastic neutron scattering in polycrystalline \Cr.
(a) The observed result.
(b) The calculated result based on the extended spin-wave theory.
Experimental and calculated results are from Ref. \ref{ref:Zhu-2019}.
}
\label{fig:experiment}
\end{figure}

\Cr~is known as a $S=3/2$ spin dimer system, as we can see in the crystal structure shown in Fig. \ref{fig:dimer}.
\cite{Zhu-2019}
Inelastic neutron scattering measurements were performed with polycrystalline samples
and they observed high-energy excitation around 12 meV
in addition to the conventional spin-wave excitation below 10 meV [see Fig. \ref{fig:experiment}(a)].
The calculated intensity of the inelastic neutron scattering is also shown in Fig. \ref{fig:experiment}(b)
based on the extended spin-wave theory.
We can see that the high-energy excitations are nicely reproduced as well as the low-energy spin-wave mode.
The used exchange interaction parameters are shown in the caption of Fig. \ref{fig:dimer}.

\begin{figure}
\begin{center}
\includegraphics[width=5cm]{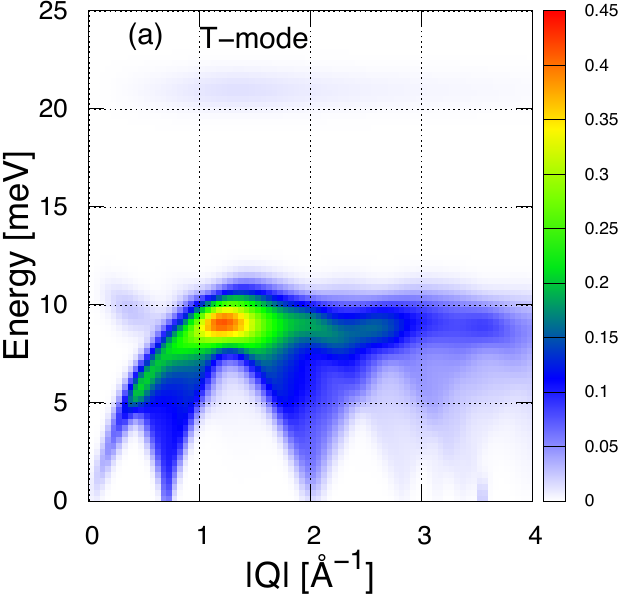}
\includegraphics[width=5cm]{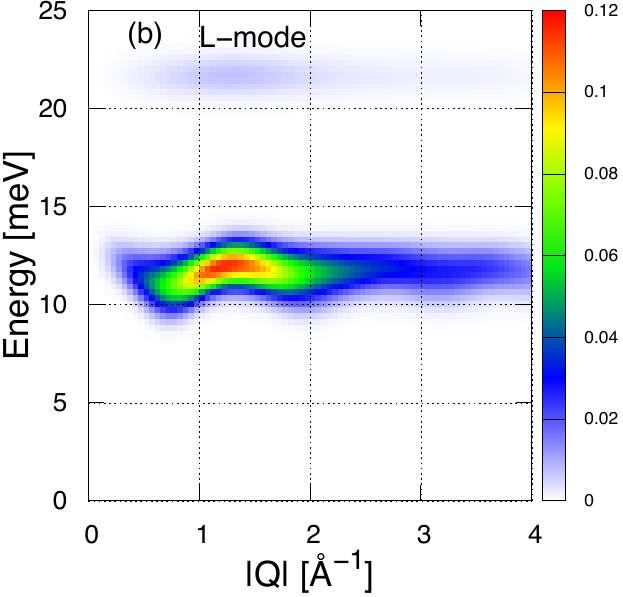}
\end{center}
\caption{
(Color online)
The calculated intensity of inelastic neutron scattering in polycrystalline \Cr.
\cite{Zhu-2019}
(a) For the T-mode.
(b) For the L-mode.
}
\label{fig:powder}
\end{figure}

\begin{figure}
\begin{center}
\includegraphics[width=7cm]{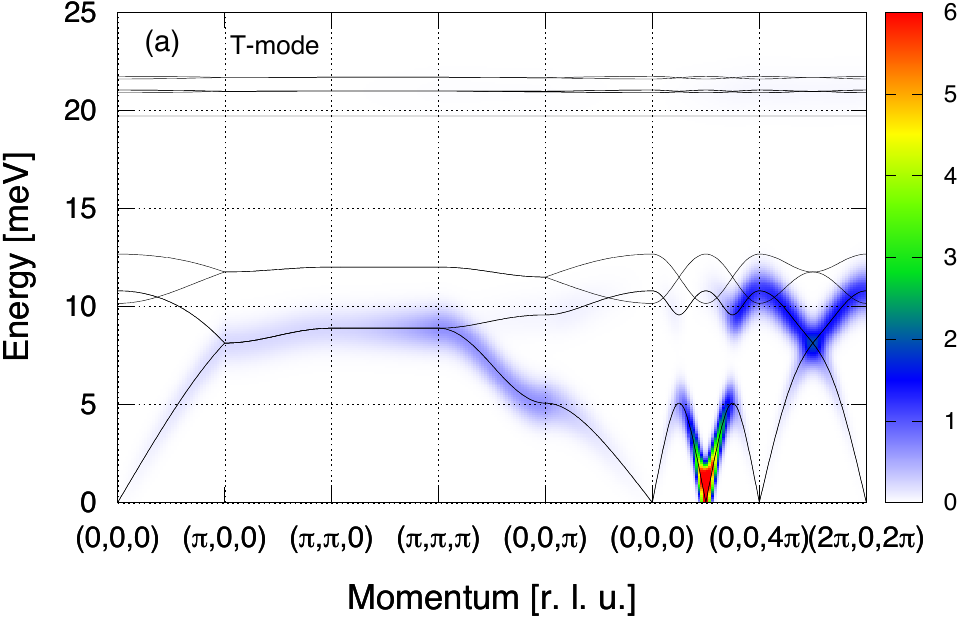}
\includegraphics[width=7cm]{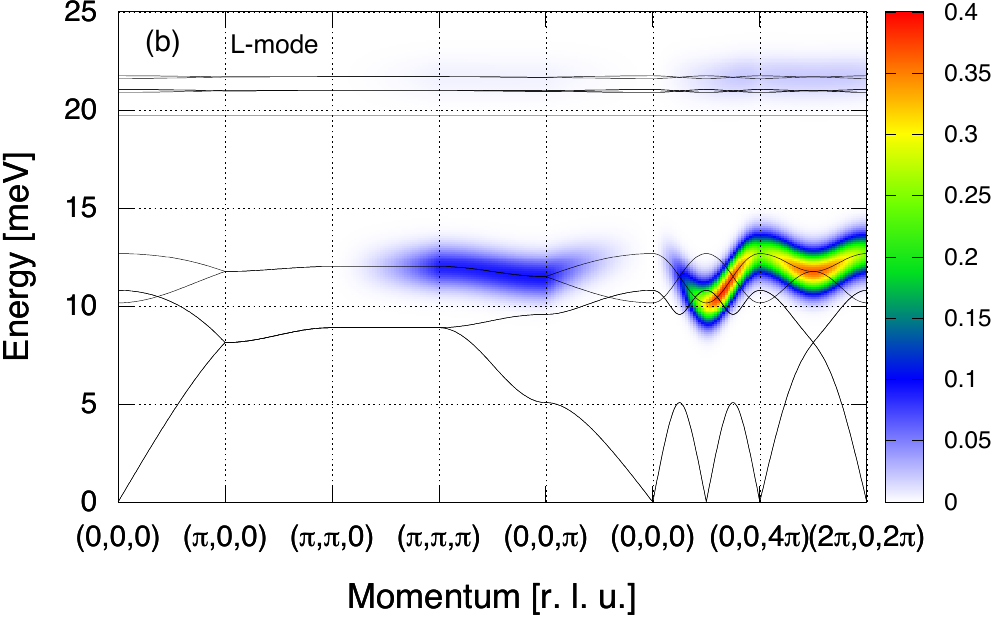}
\end{center}
\caption{
(Color online)
The calculated magnon dispersion relation and contour plot of the intensity of inelastic neutron scattering in \Cr.
\cite{Zhu-2019}
(a) For the T-mode.
(b) For the L-mode.
}
\label{fig:intensity}
\end{figure}

To understand the origin of the high-energy mode,
we resolve the intensity of the neutron scattering into transverse and longitudinal components.
As we can see in Fig. \ref{fig:powder}, the low- and high-energy modes are T- and L-modes, respectively.
In Fig. \ref{fig:intensity}, we also show the magnon dispersion relation.
The low-energy branch below 10 meV is the T-mode, while the high-energy branch above 10 meV is the L-mode.
These two excitation modes result in Fig. \ref{fig:experiment}(b) for the polycrystalline sample.
The high-energy flat modes above around 20 meV are from the high-energy spin multiplet of the $S=3/2$ dimer.
These modes have quite weak intensity as we can see in Figs. \ref{fig:experiment}(b),  \ref{fig:powder}(a), and \ref{fig:powder}(b).

On the other hand, there is another possible origin of the high-energy mode around 12 meV.
Since it is located above the conventional spin-wave mode (T-mode),
a two-magnon process of the T-mode can be the origin.
For polycrystalline samples, however, it is difficult to calculate the intensity for the two-magnon process.
Therefore, to judge the origin, we compare the integrated intensities
of the L-mode in one-magnon process and the T-mode in two-magnon process.
From Table \ref{table:S=3/2-dimer}, the former and the latter intensities are expressed as
\begin{equation}
\begin{aligned}
&I_{\rm L,1-magnon}^z = \frac{5}{4}(u^2-v^2)^2(1 - n_x - n_y), \cr
&I_{\rm T,2-magnon}^z = 5(uv)^2 ( n_x + n_y ) + \frac{1}{4}( n_x + n_y ).
\end{aligned}
\label{eqn:I2}
\end{equation}
Here, we dropped the term proportional to $4n_\rL$ in $I_{\rm T,2-magnon}$,
since it is from two-magnon process of the L-mode.
We do not consider two-magnon intensity in the $x$ and $y$ components,
since it is from a two-magnon process by the T- and L-modes.
On the basis of the extended spin-wave theory, the values of the coefficients are calculated as
$u=0.802$, $v=0.580$, and $n_x=n_y=0.0138$.
These lead to the following ratio:
\begin{align}
\frac{I_{\rm T,2-magnon}^z}{I_{\rm L,1-magnon}^z} \simeq 0.32
\end{align}
Thus, the integrated intensity of the two-magnon continuum is about 1/3 of that of the one-magnon L-mode.
The L-mode has weak dispersion and the associated magnon band is concentrated in a narrow energy region (10-12.5 meV).
In contrast, the T-mode has a wide magnon band
and we can expect a wide energy distribution (0-20 meV) for the two-magnon excitation.
Therefore, we can judge that the observed high-energy mode near 12 meV
in \Cr~stems from the longitudinal mode in the one-magnon process.

From the results in Tables. \ref{table:S=1/2-dimer} and \ref{table:S=3/2-dimer},
we can discuss a merit of large spin $S$ to observe the L-mode in spin dimer systems.
Comparing the integrated intensities in the $S=1/2$ and $S=3/2$ cases,
we notice that $I_{\rm L,1-magnon}^z$ in Eq. (\ref{eqn:I2}) is enhanced by a factor of $5$ in the $S=3/2$ case.
For $I_{\rm L,2-magnon}^z$, only the first term in Eq. (\ref{eqn:I2}) is enhanced.
Thus, $I_{\rm L,1-magnon}^z$ is relatively enhanced than $I_{\rm L,2-magnon}^z$ for large $S$.
This means that larger spin sizes of the dimer, such as $S=3/2$ and $S=2$,
are advantageous for observing the L-mode in one-magnon process.

\section{Summary}

In this paper, the total moment sum rule was derived and resolved into elastic, one-magnon, and two-magnon components,
based on the extended spin-wave theory.
The theory is applicable not only to conventional spin systems but also to systems having a quantum critical point,
such as spin dimer systems (see Table \ref{table:S=1/2-dimer})
and integer spin systems with a large single-ion anisotropy of easy-plane type (see Table \ref{table:S=1-D}).
In these systems, the ordered moment can be strongly suppressed by the quantum effect in the mean-field level
and there exists an L-mode in the magnetic excitations.
With the use of the sum rule, we can estimate the integrated intensity of the L-mode
and check the possibility to observe the L-mode in one-magnon process.
It can be also applied to estimate the integrated intensity of the T-mode in two-magnon process
and we can compare it with that of the L-mode in one-magnon process,
where both processes lead to longitudinal fluctuations of the ordered moment.

We applied the theory to an $S=1/2$ spin ladder system \Cu~and estimated the integrated intensity
of the two-magnon excitation in the disordered phase.
The result is consistent with the experiment.
The theory was extended to $S=3/2$ dimer case (see Table \ref{table:S=3/2-dimer}) and applied to \Cr.
It is confirmed that the observed high-energy mode in \Cr~is from the L-mode
rather than the T-mode in two-magnon process.
\cite{Zhu-2019}
The theory also predicts that larger spin sizes are advantageous for observing the L-mode in spin dimer systems.
Thus, the total moment sum rule derived in this paper helps us analyze and understand the measured data
of inelastic neutron scattering in the vicinity of the quantum critical point.

\vspace{1mm}
{\footnotesize\paragraph{\footnotesize Acknowledgments}
The author expresses his sincere thanks to X. Ke, S. D. Mahanti, T. Hong, and M. Koga
for useful discussions on the total moment sum rule.
This work was supported by JSPS KAKENHI Grant Number 17K05516.
}

\appendix
\setcounter{equation}{0}

\section{Conventional Spin-Wave Theory}
\label{appendix:spin-wave}

Let us briefly introduce the conventional spin-wave theory.
We consider the AF Heisenberg Hamiltonian given by Eq. (\ref{eqn:H}).
Based on the Holstein-Primakoff transformation, the spin operators on the A sublattice are expressed as
\begin{align}
&S^z_i = S - a_i^\dagger a_i, \cr
&S^+_i = \sqrt{2S} \left[ 1 - a_i^\dagger a_i/(2S) \right]^{\frac{1}{2}} a_i,
\label{eqn:Sa} \\
&S^-_i = \sqrt{2S} a_i^\dagger \left[ 1 - a_i^\dagger a_i/(2S) \right]^{\frac{1}{2}}.
\nonumber
\end{align}
Here, $a_i^\dagger$ and $a_i$ are creation and annihilation Bose operators at the $i$th site.
On the B sublattice, they are given by
\begin{align}
&S^z_i = - S + b_i^\dagger b_i, \cr
&S^+_i= \sqrt{2S} b_i^\dagger \left[ 1 - b_i^\dagger b_i/(2S) \right]^{\frac{1}{2}},
\label{eqn:Sb} \\
&S^-_i = \sqrt{2S} \left[ 1 - b_i^\dagger b_i/(2S) \right]^{\frac{1}{2}} b_i,
\nonumber
\end{align}
with $b_i^\dagger$ and $b_i$ Bose operators at the $i$th site.
We substitute Eqs. (\ref{eqn:Sa}) and (\ref{eqn:Sb}) into Eq. (\ref{eqn:H}) and use the Fourier transformation:
\begin{align}
a_i = \frac{1}{\sqrt{N/2}} \sum_\bk e^{\bk\cdot\br_i} a_\bk,~~~~~~
b_i = \frac{1}{\sqrt{N/2}} \sum_\bk e^{\bk\cdot\br_i} b_\bk.
\label{eqn:Fourier}
\end{align}
Here, $N/2$ represents number of spin sites for each A and B sublattices.
Up to the quadratic order of the Bose operator, we obtain the following Hamiltonian for the linear spin-wave theory:
\begin{align}
\H = \sum_\bk \left[ \gamma_0 ( a_\bk^\dagger a_\bk + b_\bk^\dagger b_\bk ) + \gamma_\bk ( a_\bk b_{-\bk} + a_\bk^\dagger b_{-\bk}^\dagger ) \right].
\label{eqn:H2}
\end{align}
Here, we dropped a constant term.
Reflecting the simple cubic lattice, $\gamma_\bk$ and $\gamma_0$ are given by
\begin{align}
\gamma_\bk = 2SJ ( \cos{k_x} + \cos{k_y} + \cos{k_z} ),~~~~~~
\gamma_0 = 6SJ.
\end{align}
Using the following Bogoliubov transformation,
\begin{align}
a_\bk = u_\bk \alpha_\bk - v_\bk \beta_{-\bk}^\dagger,~~~~~~
b_\bk = u_\bk \beta_\bk - v_\bk \alpha_{-\bk}^\dagger,
\label{eqn:Bogoliubov}
\end{align}
we can diagonalize the Hamiltonian in Eq. (\ref{eqn:H2}) as
\begin{align}
\H = \sum_\bk E_\bk ( \alpha_\bk^\dagger \alpha_\bk + \beta_\bk^\dagger \beta_\bk ).
\end{align}
Here, $\alpha_\bk$ and $\beta_\bk$ are bosons for the spin-wave excitation modes and we dropped a constant term.
The dispersion relation of the excitation mode is given by
\begin{align}
E_\bk
&= \sqrt{ \gamma_0^2 - \gamma_\bk^2} \cr
&= 2SJ \sqrt{ 3^2 - ( \cos{k_x} + \cos{k_y} + \cos{k_z} )^2 }.
\end{align}
The coefficients for the Bogoliubov transformation in Eq. (\ref{eqn:Bogoliubov}) are given by
\begin{align}
u_\bk = \sqrt{ \frac{1}{2} \left( \frac{\gamma_0}{E_\bk} + 1 \right) },~~~~~~
v_\bk = \sqrt{ \frac{1}{2} \left( \frac{\gamma_0}{E_\bk} - 1 \right) } \frac{\gamma_\bk}{|\gamma_\bk|}.
\label{eqn:uv}
\end{align}



\begin{thebibliography}{99}
\bibitem{note:sz}
Using the matrix form in Eq. (\ref{eqn:sz-matrix}), we can express $\Delta S^z$ as
$\Delta S^z=0|1\rangle\langle 1| + 1|0\rangle\langle 0| + 2|-1\rangle\langle -1|$.
Then, we obtain
$
(\Delta S^z)^2
=1^2|0\rangle\braket{0|0}\langle 0| + 2^2|-1\rangle\braket{-1|-1}\langle -1|
=|0\rangle\langle 0| + 4|-1\rangle\langle -1|
$.
\end{thebibliography}
\end{document}